\documentclass[journal]{IEEEtran}
\ifCLASSINFOpdf
\else
\fi

\usepackage{booktabs}
\usepackage{graphicx}
\usepackage{amsfonts}
\usepackage{eucal}
\usepackage{amsmath}
\usepackage{multirow}
\usepackage{subfigure}
\usepackage[ruled, lined, vlined]{algorithm2e}

\usepackage{booktabs}
\usepackage{listings}
\usepackage{enumitem}
\usepackage{algpseudocode}
\usepackage{graphicx}
\usepackage{color}
\usepackage{threeparttable}
\usepackage{setspace}
\usepackage[small]{caption}
\usepackage{url}
\usepackage{ragged2e}
\usepackage{xcolor}
\begin{document}
%
\title{DeepFM: An End-to-End Wide \& Deep Learning Framework for CTR Prediction}
%
%
%
\author{Huifeng Guo,
        Ruiming Tang,
        Yunming Ye,
        Zhenguo Li,
        Xiuqiang He,
        and~Zhenhua Dong
\IEEEcompsocitemizethanks{
\IEEEcompsocthanksitem This work is done when Huifeng Guo worked as intern at Noah's Ark Research Lab of Huawei.
\IEEEcompsocthanksitem Yunming Ye is the corresponding author.
\IEEEcompsocthanksitem Huifeng Guo and Yunmming Ye are with Shenzhen Key Laboratory of Internet Information Collaboration, Shenzhen Graduate School, Harbin Institute of Technology, Shenzhen, 518055, P.R.China.\protect\\
E-mail: huifengguo@yeah.net, yeyunming@hit.edu.cn
\IEEEcompsocthanksitem Ruiming Tang, Xiuqiang He and Zhenhua Dong are with Noah's Ark Research Lab of Huawei, Shenzhen, P.R.China.\protect\\
E-mail: \{tangruiming, hexiuqiang, dongzhenhua\}@huawei.com
\IEEEcompsocthanksitem Zhenguo Li is with Noah's Ark Research Lab of Huawei, Hong Kong, P.R.China.\protect\\
E-mail: li.zhenguo@huawei.com}
}
\markboth{Journal of \LaTeX\ Class Files,~Vol.~14, No.~8, August~2015}%
{Shell \MakeLowercase{\textit{et al.}}: Bare Demo of IEEEtran.cls for IEEE Journals}
%



\maketitle

\begin{abstract}

  Learning sophisticated feature interactions behind user behaviors is critical in maximizing CTR for recommender systems. Despite great progress, existing methods have a strong bias towards low- or high-order interactions, or rely on expertise feature engineering. In this paper, we show that it is possible to derive an end-to-end learning model that emphasizes both low- and high-order feature interactions. The proposed framework, DeepFM, combines the power of factorization machines for recommendation and deep learning for feature learning in a new neural network architecture. Compared to the latest Wide \& Deep model from Google, DeepFM has a shared raw feature input to both its ``wide'' and ``deep'' components, with no need of feature engineering besides raw features. DeepFM, as a general learning framework, can incorporate various network architectures in its deep component. In this paper, we study two instances of DeepFM where its ``deep" component is DNN and PNN respectively, for which we denote as DeepFM-D and DeepFM-P. Comprehensive experiments are conducted to demonstrate the effectiveness of DeepFM-D and DeepFM-P over the existing models for CTR prediction, on both benchmark data and commercial data. We conduct online A/B test in Huawei App Market, which reveals that DeepFM-D leads to more than 10\% improvement of click-through rate in the production environment, compared to a well-engineered LR model. We also covered related practice in deploying our framework in Huawei App Market.

%
\end{abstract}


%
\IEEEpeerreviewmaketitle

\section{Introduction}\label{section:intro}

\IEEEPARstart{W}{ith} the rapid development of the Internet and mobile devices, our daily activities connect closely to online services, such as online shopping, online news and videos, online social networks, and many more. Recommender systems are a powerful information filter for guiding users to find the items of interest in a gigantic and rapidly expanding pool of candidates. As elaborated in~\cite{alpha-NDCG}, if an IR system's response to each query is a ranking of documents in decreasing order of probability of relevance, the overall effectiveness of the system to its user will be maximized. With this principle, the prediction of click-through rate (CTR) is crucial for recommender systems, where the task is to estimate the probability a user will click on a recommended item. In many recommender systems, the goal is to maximize the number of clicks, and the items returned to a user are ranked by the estimated CTR; while in other application scenarios such as online advertising it is also important to improve revenue, so the ranking strategy can be adjusted accordingly, such as by CTR$\times$bid with ``bid'' being the benefit the system receives once the item is clicked. In either case, the key is in estimating CTR precisely.

\begin{figure*}[ht]
\centering
\includegraphics[width=1\textwidth]{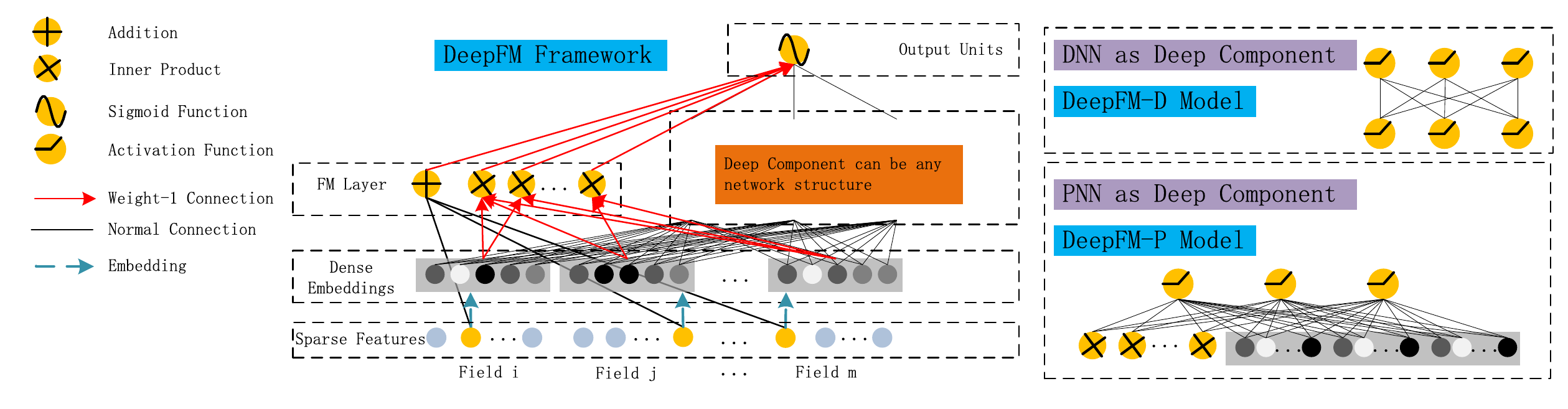}
\caption{{Wide \& deep architecture of the DeepFM framework (left). The wide and deep component share the same input raw feature vector, which enables DeepFM to learn low- and high-order feature interactions simultaneously from the input raw features. The wide component of DeepFM is an FM layer, which we refer to as \emph{FM Component} in the paper. The \emph{Deep Component} of DeepFM can be any neural network. In this paper, we will study two instances of DeepFM, namely DeepFM-D(top-right) and DeepFM-P (lower-right), the deep components of which are DNN and PNN respectively.}}\label{fig:architecture}
\end{figure*}

It is important for CTR prediction to learn implicit feature interactions behind user click behaviors. By our study in a mainstream apps market, we found that people often download apps for food delivery at meal-time, suggesting that the (order-2) interaction between app category and time-stamp can be used as a signal for CTR prediction. As a second observation, male teenagers like shooting games and RPG games, which means that the (order-3) interaction of app category, user gender and age is another signal for CTR prediction. In general, such interactions of features behind user click behaviors can be highly sophisticated, where both low- and high-order feature interactions should play important roles. According to the insights of the Wide \& Deep model~\cite{wide-n-deep} from Google, considering low- and high-order feature interactions \emph{simultaneously} brings additional improvement over the cases of considering either alone.

The key challenge is in effectively modeling feature interactions. Some feature interactions can be easily understood and engineered by experts (like the instances above). However, most other feature interactions are hidden in data and difficult to identify \emph{a priori} (for instance, the classic association rule ``diaper and beer" is mined from a large amount of data, instead of being discovered by experts), which can only be captured \emph{automatically} by machine learning. Even for easy-to-understand interactions, it seems unlikely for experts to model them exhaustively, especially when the number of raw features is huge.

Despite their simplicity, generalized linear models, such as \emph{FTRL}~\cite{FTRL}, have shown decent performance in practice. However, a (generalized) linear model lacks the ability to learn feature interactions, and a common practice is to manually include pairwise feature interactions in designing its feature vector. Such a method is hard to generalize to model high-order feature interactions or those never or rarely appear in the training data~\cite{fm}. \emph{Factorization Machines (FM)}~\cite{fm} model pairwise feature interactions as inner product of latent vectors between features and show very promising results. While in principle FM can model high-order feature interactions, in practice usually only order-2 feature interactions are considered due to high complexity.

As a powerful approach to learning feature representations, deep neural networks have shown the potential to learn sophisticated feature interactions automatically. Some ideas extend CNN and RNN for CTR prediction \cite{cnn,rnn}, but CNN-based models are biased to the interactions between neighboring features while RNN-based models are more suitable for click data with sequential dependency. 
\cite{fnn} studies feature representations and proposes \emph{Factorization-machine supported Neural Network (FNN)}. This model pre-trains FM before applying DNN, thus limited by the capability of FM. Feature interaction is studied in \cite{pnn}, by introducing a product layer between embedding layer and fully-connected layers, and proposing the \emph{Product-based Neural Network} (\emph{PNN}). As noted in~\cite{wide-n-deep}, PNN and FNN, like other deep models, capture little low-order feature interactions, which are also essential for CTR prediction. To model both low- and high-order feature interactions, \cite{wide-n-deep} proposes an hybrid network structure (\emph{Wide \& Deep}) that combines a linear (``wide") model and a deep model. In this hybrid model, two different inputs are required for the ``wide" part and ``deep" part, respectively, and the input of ``wide" part still relies on feature engineering from domain experts.

One can see that existing models are biased to low- or high-order feature interactions, or rely on feature engineering. In this paper, we develop a learning framework that is able to learn feature interactions of all orders in an end-to-end manner, without any feature engineering besides raw features. Our main contributions are summarized as follows:

\begin{itemize}[leftmargin=*]
\item We propose a new neural network framework DeepFM (left part of Figure~\ref{fig:architecture}) that integrates the architectures of FM and deep neural networks in a way that models low-order feature interactions like FM and models high-order feature interactions like deep neural networks. Unlike the wide \& deep model~\cite{wide-n-deep}, DeepFM can be trained end-to-end without any feature engineering. There is no implicit requirement on the network structure of the ``deep" part of DeepFM, so various deep architectures are plausible\footnote{In all figures of this paper, a \textbf{Normal Connection} in black refers to a connection with weight to be learned; a \textbf{Weight-1 Connection}, red arrow, is a connection with weight 1 by default; \textbf{Embedding}, blue dashed arrow, means a latent vector to be learned; \textbf{Addition} means adding all input together; \textbf{Product}, including \textbf{Inner-} and \textbf{Outer-Product}, means the output of this unit is the product of two input vector; \textbf{Sigmoid Function} is used as the output function in CTR prediction; \textbf{Activation Functions}, such as relu and tanh, are used for non-linearly transforming the signal;The yellow and blue circles in the sparse features layer represent one and zero in one-hot encoding of the input, respectively.}.
\item DeepFM can be trained efficiently because its ``wide" part and ``deep" part share the same input and also the embedding vector. In contrast, the input vector to Google's Wide \& Deep model~\cite{wide-n-deep} can be of huge size as it includes manually designed pairwise feature interactions in its wide part, which also greatly increases its complexity. Moreover, no expertise feature engineering is needed in our DeepFM framework, while it is required in~\cite{wide-n-deep}, as the performance of linear models relies heavily on feature engineering. We study two instances of DeepFM in detail, namely DeepFM-D and DeepFM-P (see the right part of Figure~\ref{fig:architecture}), where the ``deep" part of the DeepFM framework is DNN and PNN, respectively.
\item We evaluate DeepFM-D and DeepFM-P on both benchmark data and commercial data, which shows consistent improvement over existing models for CTR prediction.
\item To investigate whether DeepFM is well suited in a production environment as a further exploration of our previous work~\cite{HFGuo17DeepFM}, we conduct the online A/B test in
Huawei App Market. The results  reveal that DeepFM-D model leads to more than 10\% improvement of CTR, compared to a well-engineered LR model. We also covered our practice in deploying our framework, such as multi-GPU data parallelism and asynchronous data reading. Extensive experiments are conducted to show the effectiveness of our proposed techniques.
\end{itemize}

The rest of the paper is organized as follows. Section~\ref{section:related} gives an overview of some related works on recommender systems. Section~\ref{section:App} presents our DeepFM framework, as well as two instances DeepFM-D and DeepFM-P of our framework. In Section~\ref{section:exp}, We extensively evaluate the effectiveness and efficiency of DeepFM-D and DeepFM-P on benchmark datasets and commercial dataset, and we conduct online A/B test in Huawei App Market. Finally, we conclude our work in Section~\ref{section:conclusion}.

\section{Related Work}\label{section:related}

%

In this section, we review three important categories of models in recommender systems, namely \emph{collaborative filtering models}, \emph{linear models} and \emph{deep learning models}.

\subsection{Collaborative Filtering in Recommender Systems}

The collaborative filtering (CF)-based models are well studied for recommender systems from the last decade. The basic assumption is that users with similar past behaviors will like the same kind of items, and the items attracting similar users will share similar ratings from a user. Collaborative filtering models consist of memory-based and model-based methods. The memory-based methods directly define the similarities between pairs of users and pairs of items, which are used to calculate the ratings of unknown user-item pairs \cite{item-cf}. Although it is easy to implement and explain, there are several limitations for memory-based methods. For instance, the similarity values are unreliable when the data is sparse and the common items are few. As a complementary part, model-based methods define a model to fit the known user-item interactions and predict the rating of unknown user-item pairs using the learned model. The most widely used model-based model is matrix factorization (MF)~\cite{mf,svd++}. Based on the low rank assumption and the observed user-item interactions, MF models characterize both items and users by vectors in the same space and predict an unknown rating of a user-item pair relies on the item and user vectors.

Different from CF-based models, the content-based models rely on user portrait or product information \cite{contentbased-1}. However, both CF-based and  content-based models have limitations. While the former does not explicitly incorporate  the users' and items' feature information, the latter does not necessarily consider the information in preference similarity across individuals. Therefore hybrid methods have gained popularity in recent years. In order to incorporate user-item interaction information and auxiliary information together, such as text, temporal information, location information and so on, several hybrid methods \cite{ColTopicRec11,svdfeature,GeoFM} are proposed. By learning side information and user-item interaction simultaneously, hybrid methods are able to alleviate cold-start problem and give the better recommendation.

Above models are not applied in CTR prediction very often, for reasons such as poor scalability, unsatisfactory performance on sparse data.

\subsection{Linear Models in Recommender Systems}

Because of the robustness and efficiency, generalized Logistic Regression (LR) models, such as FTRL~\cite{FTRL}, are widely used in CTR prediction. To learn feature interactions, a common practice is to manually include pairwise feature interactions in its feature vector. The Poly-2~\cite{Poly2SVM10} model is proposed to model all order-2 feature interactions to avoid feature engineering. Factorization Machines (FM)~\cite{fm} adopts a factorization technique to improve the ability of learning feature interactions when data is very sparse. Recently, to model interaction features from different fields, the authors of FFM~\cite{ffm} introduces field information into FM model. LR, Poly-2 and FM variants are widely used in CTR prediction in industry. In addition, a few other models are also proposed for CTR prediction, such as LR+GBDT model~\cite{facebookgbdt}, tensor based model~\cite{PITFRendleS10}, and bayesian model~\cite{bayesCTR10}.

\subsection{Deep Learning in Recommender Systems}
Due to the powerful ability of feature learning, deep learning models have achieved great success in various areas, such as computer vision \cite{residual2016}, natural language processing \cite{Word2vectorMikolov13}, audio recognition \cite{RNNforSpeech} and gaming~\cite{AlphaGoSilverHMGSDSAPL16}. In order to take advantage of its feature learning ability to enhance recommender systems, several deep learning models are proposed for recommendation (e.g., \cite{youtube,other-2,other-7,other-6, wsdm17recurentRS, rnn, jointdeepforreviewZhengNY17, pnn, fnn, cnn}).
These works can be divided into CTR-based and rating-based models.

CTR-based models \cite{pnn, fnn, cnn, rnn} are already mentioned in Section~\ref{section:intro} and some of them will be discussed again in Section~\ref{section:App}, therefore we omit them here.

There are two kinds of rating-based deep models, CF-based and hybrid models. The CF-based models, such as \cite{other-2,other-3,other-4,other-6},  are proposed to improve Collaborative Filtering via deep learning. For instance, \cite{other-3} and \cite{other-2} complete the rating matrix by auto encoder~\cite{SDA2010} and restricted boltzmann machine, respectively. Unlike CF-based models, the hybrid-based models \cite{other-1,other-7, jointdeepforreviewZhengNY17} use deep learning to learn features of various domains. Specifically, \cite{wsdm17recurentRS} proposes a recurrent recommender system, which is able to capture temporal information and predict future behavioral trajectories. \cite{jointdeepforreviewZhengNY17} utilizes both review information and user-item interactions. For the purpose of learning better features, \cite{other-1} designs a novel end-to-end model to learn features from audio content and user-item interactions simultaneously to make personalized recommendations. In order to ease the cold start problem when recommending new and unpopular songs, \cite{other-7} adds deep convolution neural network in the latent factor framework to learn audio features better.

Several models are proposed in industry. Google develops a two-stage deep learning framework for YouTube video recommendation \cite{youtube}. In order to learn the relationship between image features and other features, Alibaba proposes an end-to-end deep model \cite{cnn_image}, which incorporates convolutional neural network for learning image features and multi-layer perception for other features, for CTR prediction.
Readers interested in deep learning models in recommender systems can refer to a comprehensive survey work~\cite{deep-rec-survey}.

\section{Our Approach}\label{section:App}


Suppose the training data consists of $n$ instances $(\chi,y)$, where $\chi$ is an $m$-fields data record usually involving a pair of user and item, and $y\in\{0,1\}$ is the associated label indicating user click behaviors ($y=1$ means the user clicked the item, and $y=0$ otherwise). $\chi$ may include categorical fields (e.g., gender, location) and continuous fields (e.g., age). Each categorical field is represented as a one-hot vector, and each continuous field is represented as the value itself, or a one-hot vector after discretization. Thus, each instance is converted to $(x,y)$ where $x=[x_{\rm field_1},x_{\rm field_2}, ...,x_{\rm filed_j},...,x_{\rm field_m}]$ is a $\rm d$-dimensional vector, with $x_{\rm field_j}$ being the vector representation of the $\rm j^{th}$ field of $\chi$. Normally, $x$ is high-dimensional and extremely sparse. The task of CTR prediction is to develop a prediction model to estimate the probability of a user clicking a specific item in a given context.

\subsection{DeepFM}\label{section:App:model}

To learn both low- and high-order feature interactions, we propose an end-to-end deep learning framework for CTR prediction, namely Factorization-Machine based neural network (DeepFM). As depicted in Figure~\ref{fig:architecture}, DeepFM consists of two components, \emph{FM Component} and \emph{Deep Component}, that share  the same input. For feature ${\rm i}$, a scalar $w_{\rm i}$ is used to weigh its order-1 importance, a latent vector $V_{\rm i}$ is used to measure its impact of interactions with other features. $V_{\rm i}$ is fed in FM component to model order-2 feature interactions, and fed in deep component to model high-order feature interactions. All parameters, including $w_{\rm i}$, $V_{\rm i}$, and the network parameters ($W^{(\rm l)}$, $b^{(\rm l)}$ below) are trained jointly for the combined prediction model:

\begin{equation}
\hat{y}(x)=sigmoid(y_{\rm FM}(x)+y_{\rm Deep}(x))
\label{eq:DeepFMCTR}
\end{equation}
where $\hat{y}(x)\in (0,1)$ is the predicted CTR, $y_{\rm FM}(x)$ is the output of FM component, and $y_{\rm Deep}(x)$ is the output of deep component. We also present two instances of DeepFM framework in Figure~\ref{fig:architecture}, namely DeepFM-D and DeepFM-P, whose deep component is DNN and PNN, respectively. The prediction formulae of DeepFM-D and DeepFM-P update Equation~\ref{eq:DeepFMCTR} by setting $y_{\rm Deep}(x) = y_{\rm DNN}(x)$ and $y_{\rm Deep}(x) = y_{\rm PNN}(x)$, respectively. The definitions of $y_{\rm FM}(x)$, $y_{\rm DNN}(x)$ and $y_{\rm PNN}(x)$ are introduced in the following sections.

\subsubsection{FM Component of DeepFM}\label{section:App:model:fm}

\begin{figure}[ht]
\centering
\includegraphics[width=0.48\textwidth]{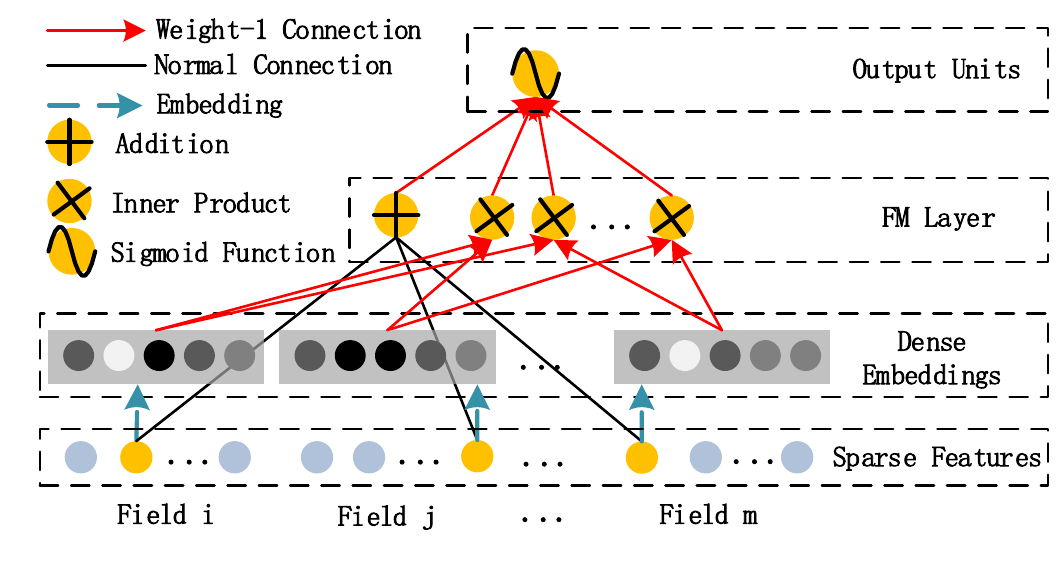}
\caption{The architecture of FM. }\label{fig:fm}
\end{figure}

The FM component is a factorization machines, which is proposed in \cite{fm} to learn feature interactions for recommendation. Besides a linear (order-1) interactions among features, FM models pairwise (order-2) feature interactions as inner product of respective feature latent vectors.
It can capture order-2 feature interactions much more effectively than previous approaches (such as LR and Poly-2~\cite{Poly2SVM10}) especially when the data is sparse. For comparison, we show the prediction models for LR and Poly-2:

\begin{equation}
\centering
y_{\rm LR}(x)=\left< w,x  \right> +\sum_{\rm i,j}w_{\rm h(i,j)}  x_{\rm i}\cdot x_{\rm j},
\label{eq:LR-model}
\end{equation}

\begin{equation}
\centering
y_{\rm Poly-2}(x)=\left< w,x  \right> +\sum_{\rm i=1}^{\rm d}\sum_{\rm j=i+1}^{\rm d}w_{\rm h(i,j)}  x_{\rm i}\cdot x_{\rm j}
\label{eq:poly2-model}
\end{equation}
where $y_{\rm LR}$ and $y_{\rm Poly-2}$ are the prediction of LR model and Poly-2 models, and ${\rm h(i,j)}$ is a function encoding ${\rm i}$ and ${\rm j}$ into a natural number. LR models linear combination of the features and some order-2 feature interactions that are selected by experts (i.e., $\rm i$ and $\rm j$ are picked by human). To avoid feature engineering, Poly-2 chooses to model all possible order-2 feature interactions. In these two approaches, each feature interaction is assigned with a parameter, so that the number of parameters in the model is huge. Moreover, the parameter of an interaction of features ${\rm i}$ and ${\rm j}$ can be learned only when feature $\rm i$ and feature $\rm j$ both appear in a sample.

While in FM, it is measured via the inner product of their latent vectors $V_{\rm i}$ and $V_{\rm j}$. Thanks to this flexible design, FM can train latent vector $V_{\rm i}$ ($V_{\rm j}$) whenever $\rm i$ (or $\rm j$) appears in a data record. Therefore, feature interactions, which are never or rarely appeared in the training data, are better learned by FM. As Figure~\ref{fig:fm} shows, the output of FM is the summation of an \textbf{Addition} unit and a number of \textbf{Inner Product} units:
\begin{equation}
\centering
y_{\rm FM}(x)=\left< w,x  \right> +\sum_{\rm i=1}^{\rm d}\sum_{\rm j=i+1}^{\rm d}\left< V_{\rm i},V_{\rm j}  \right>  x_{\rm i}\cdot x_{\rm j},
\label{eq:FM-model}
\end{equation}
where $w \in R^{\rm d}$ and $V_{\rm i}\in R^{\rm k}$ ($\rm k$ is given)\footnote{We omit a constant offset for simplicity.}. The Addition unit ($\left< w,x  \right>$) reflects the importance of order-1 features, and the Inner Product units represent the impact of order-2 feature interactions. As presented in Equation~\ref{eq:DeepFMCTR}, the output of FM component $y_{\rm FM}(x)$ is part of the final CTR prediction.

\subsubsection{Deep Component of DeepFM}\label{section:App:model:dnn}

The deep component is a feed-forward neural network, which is used to learn high-order feature interactions. A data record (a vector) is fed into the neural network. Compared to neural networks with image~\cite{residual2016} or audio~\cite{RNNforSpeech} data as input, which is purely continuous and dense, the input of CTR prediction is quite different, which requires a new network architecture design. Specifically, the raw feature input vector for CTR prediction is usually highly sparse\footnote{Only one entry is non-zero for each field vector.}, super high-dimensional\footnote{E.g., in an app store of billion users, the one field vector for user ID is already of billion dimensions.}, categorical-continuous-mixed, and grouped in fields (e.g., gender, location, age). This suggests an embedding layer to compress the input vector to a low-dimensional, dense real-value vector before further feeding into the first hidden layer, otherwise the network can be overwhelming to train.
\begin{figure}[ht]
\begin{center}
\includegraphics[width=0.48\textwidth]{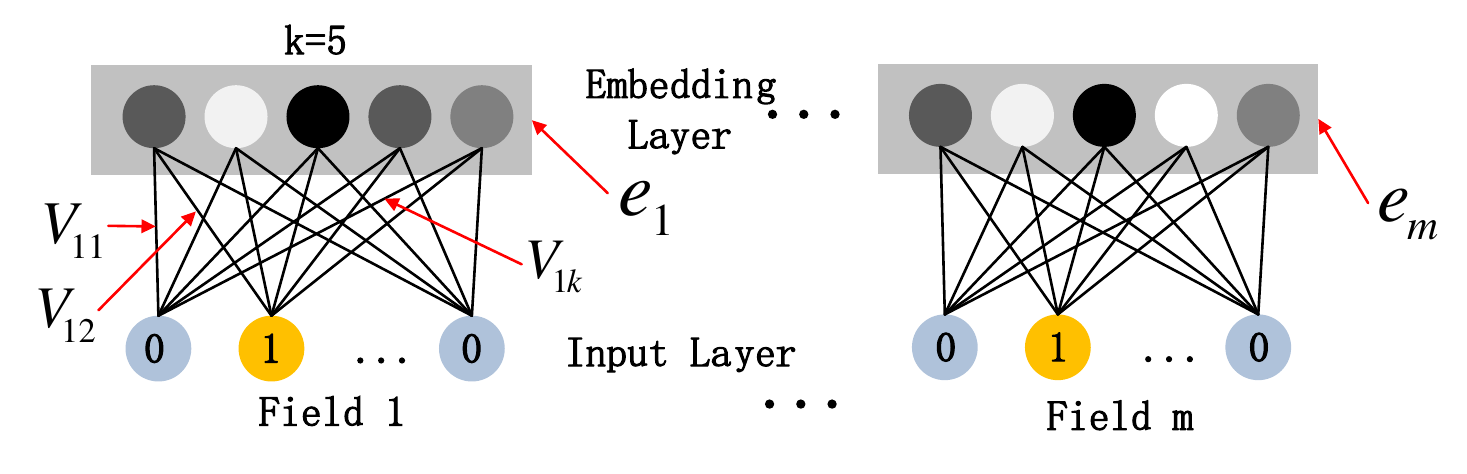}
\caption{The structure of the embedding layer.}\label{fig:embed}
\end{center}
\end{figure}

Figure \ref{fig:embed} highlights the network structure from the input layer to the embedding layer. We would like to point out two interesting designs of this network structure: 1) while input field vectors can be of different sizes, their embeddings are of the same size ($k$); 2) the latent feature vectors ($V$) in FM now serve as network weights which are learned and used to compress the input field vectors to the embedding vectors. In \cite{fnn}, $V$ is pre-trained by FM and used as initialization. In this work, rather than using the latent feature vectors of FM to initialize the networks as in \cite{fnn}, we include the FM model as part of our overall learning architecture. As such, we eliminate the need of pre-training by FM and instead jointly train the entire network in an end-to-end manner.

Denote the output of the embedding layer as:
\begin{equation}
\label{eq:embed}
a^{(\rm 0)}=[e_{\rm 1},e_{\rm 2},...,e_{\rm m}],
\end{equation}
\begin{equation}
\label{eq:embed}
e_{\rm i}= V_{\rm field_{i}}\cdot x_{\rm field_{i}},
\end{equation}
where $e_{\rm i}$ is the embedding of the $\rm i$-th field, $\rm m$ is the number of fields, $V_{\rm field_{i}}$ is the parameters between the embedding layer and the input layer of the ${\rm i}^{\rm th}$ field (as shown in Figure~\ref{fig:embed}), $x_{\rm field_{\rm i}}$ is the one-hot vector of the ${\rm i}^{\rm th}$ field raw input data.

It is worth pointing out that FM component and deep component share the same feature embedding, which brings two important benefits: 1) it learns both low- and high-order feature interactions from raw features; 2) there is no need for expertise feature engineering of the input, as required in Google Wide \& Deep model~\cite{wide-n-deep}.

Note that in the proposed DeepFM framework, there is no implicit requirement on the network structure of the deep component. In this section, we show only a general deep component of DeepFM. In the next sections, we will present in detail the network structure of the deep component of two instances of the DeepFM framework, called DeepFM-D and DeepFM-P.

\subsubsection{Deep Component of DeepFM-D Model}

\begin{figure}[ht]
\centering
\includegraphics[width=0.48\textwidth]{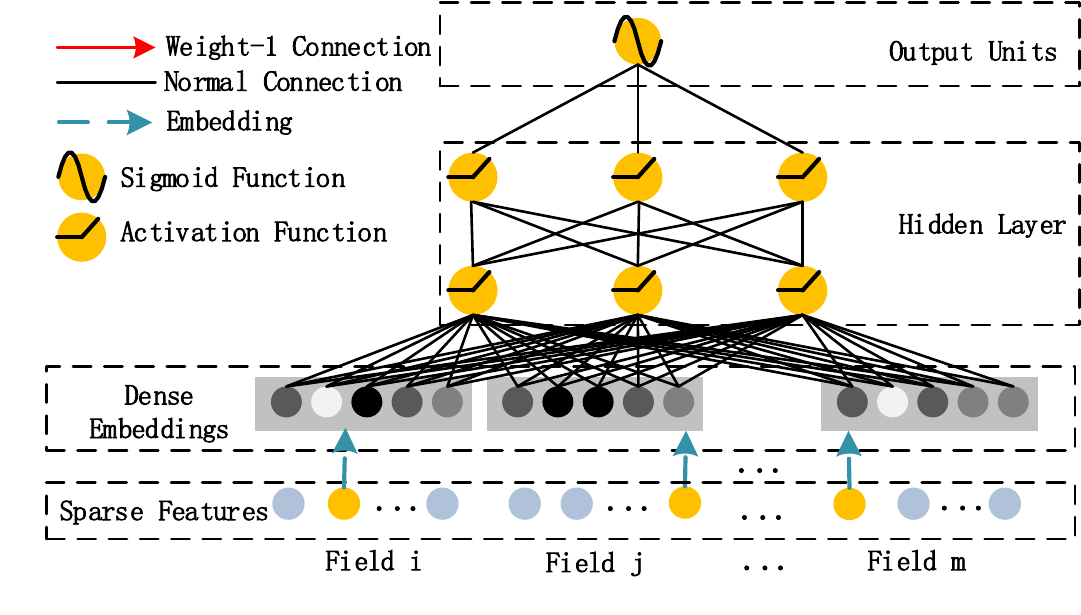}
\caption{The architecture of the deep component of DeepFM-D.}\label{fig:dnn}
\end{figure}

The deep component of DeepFM-D is a fully-connected deep neural network (DNN, or equivalently Multilayer Perceptron). The structure of DNN is presented in Figure~\ref{fig:dnn}.

In such network structure, output of the embedding layer $a^{(0)}$ is fed into the deep neural network, and the forward process is:

\begin{equation}
\centering
\label{eq:h2h}
a^{\rm (l+1)} = \sigma(W^{\rm (l)}\cdot a^{\rm (l)} + b^{\rm (l)}),
\end{equation}
where $\rm l$ is the layer depth and $\sigma$ is an activation function. $a^{\rm (l)}$, $W^{\rm (l)}$, $b^{\rm (l)}$ are the output, model weight, and bias of the $\rm l$-th layer, respectively. After going through several hidden layers, a dense real-value feature vector is generated as,
\begin{equation}
\centering
\label{eq:dnn}
y_{\rm DNN}(x)={W^{\rm |H|+1}\cdot a^{\rm |H|}+b^{\rm |H|+1}},
\end{equation}
where $\rm |H|$ is the number of hidden layers. This feature vector $y_{\rm DNN}(x)$ is finally fed into the sigmoid function for CTR prediction, as described in Equation~\ref{eq:DeepFMCTR}.

\subsubsection{Deep Component of DeepFM-P Model}

The deep component of DeepFM-P is a product based neural network (PNN)~\cite{pnn}. As presented in~\cite{pnn}, there are three variants of PNN models, i.e., IPNN, OPNN, PNN$\ast$. PNN introduces a product layer between the embedding layer and the first hidden layer (the middle part of Figure~\ref{fig:othermodel}). The three variants differ from each other in defining different product operations between features as feature interactions. More specifically, IPNN uses inner product, OPNN uses outer product, and PNN$\ast$ uses both inner and outer product. The details of three PNN variants are presented as follows.

In the product layer of PNN model (the middle part of Figure~\ref{fig:othermodel}), it consists of two parts: linear neurons (right part of the layer) and quadratic neurons (``product symbol" in the left part of the layer). Linear neurons are the concentration of the embedding vectors of all fields, while quadratic neurons are the products of embedding vectors from a pair of fields.
The output unit of the first hidden layer is

\begin{equation}
\centering
\label{eq:pnn}
a^{\rm (1)} = \sigma(W^{\rm (0)}_{\rm linear}\cdot z + W^{\rm (0)}_{\rm quadratic}\cdot p + b^{\rm (0)}).
\end{equation}
In Equation~\ref{eq:pnn}, $z = a^{\rm (0)} = [e_{\rm 1},e_{\rm 2},...,e_{\rm m}]$ is the output vector of linear neurons in the product layer, which is embedding vectors of different fields themselves. $p = \{g(e_{\rm i}, e_{\rm j})\}$ ${\rm (i\in [1,m], {\rm j}\in [i+1,m], j > i)}$ is the output vector of quadratic neurons, which includes the interactions between any two embedding vectors $e_{\rm i}$ and $e_{\rm j}$. $W^{\rm (0)}_{\rm linear}, W^{\rm (0)}_{\rm quadratic}$ are parameters between the product layer and the first hidden layer connecting to linear neurons and quadratic neurons respectively. The three variants of PNN define function $g$ differently: IPNN and OPNN define $g$ to be the inner product and outer product of two vectors respectively, while PNN$\ast$ considers both inner and outer product. Finally, going through several fully-connected hidden layers (as defined in Equation~\ref{eq:h2h}), $y_{\rm PNN}(x)$ has a similar output value as $y_{\rm DNN}(x)$ (as defined in Equation~\ref{eq:dnn}).

\subsection{Practical Issues}\label{subsection:learning}

\subsubsection{Learning}
\begin{itemize}[leftmargin=1em]
\item \textbf{Objective function}: In the domain of CTR prediction, the most commonly used objective function is Logloss, which is equivalent to the K-L divergence between two distributions:

\begin{equation}
\centering
\label{eq:logloss}
e(x,y) = -(y\cdot {\rm log} (\hat{y}(x)) + (1-y)\cdot (1-{\rm log}(1-\hat{y}(x))))
\end{equation}
\begin{equation}
\centering
\label{eq:logloss}
E = \sum_{(x,y)\in T}e(x,y)
\end{equation}
where $(x,y)$ is a data instance, $x$ is the feature vector and $y$ is the label, $\hat{y}(x)$ is the prediction of the instance $x$, $e(x,y)$ is the Logloss of $(x,y)$, and $E$ is the Logloss of dataset $T$.

%

\item \textbf{Overfitting}: In machine learning, one important issue is to prevent overfitting. An overfit model has poor performance since it overreacts to the given training data. The authors of~\cite{fm_l2} state that FM may suffer from overfitting and thus they utilize L2-norm to regularize the objective function. On the other hand, as a complicated model, neural networks are also easy to overfit. The authors of~\cite{dropout14} propose a simple strategy to prevent neural networks from overfitting, which is known as dropout. Therefore, in our DeepFM framework, we adopt L2-norm to regularize the FM component and adopt dropout for the deep component.
\end{itemize}
\subsubsection{Accelerating Strategy}\label{subsection:Accelerating}
\begin{itemize}[leftmargin=1em]
\item \textbf{Multi-GPU Architecture}: In real applications, the amount of training data is so huge that the training process has to take a long time. For the purpose of accelerating this process, we utilize the multi-GPU data parallelism when deploying DeepFM in the production environment (as presented in Figrue~\ref{fig:multi-gpu}). At first, we split a batch of data records into \texttt{Num\_GPU} pieces (\texttt{Num\_GPU} is the number of GPU cards) and feed each piece into different GPU cards simultaneously. Then, the gradient of data records in different pieces is computed by individual GPU card. After that, the gradient is collected and averaged. Finally, the model parameters are updated by the averaged gradient.

\begin{figure}[ht]
\centering
\includegraphics[width=0.48\textwidth]{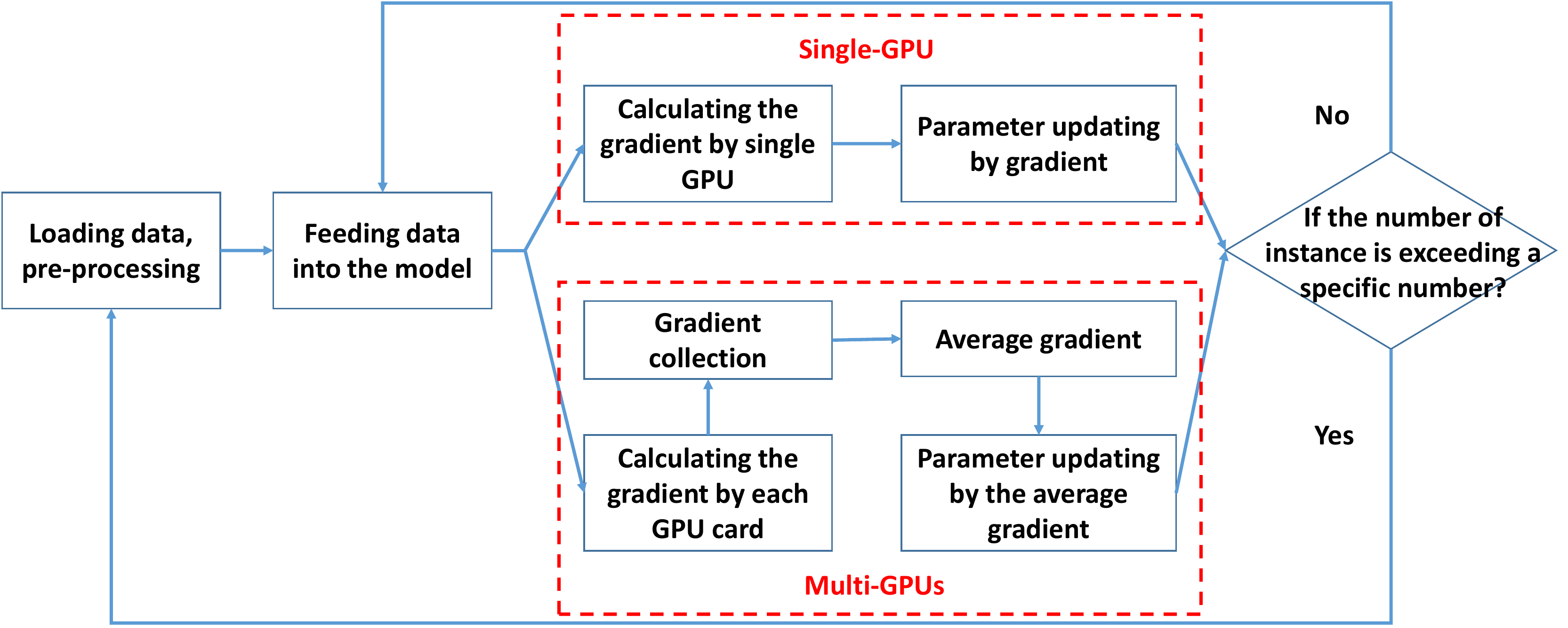}
\caption{The Multi-GPU architecture.}\label{fig:multi-gpu}
\end{figure}

\begin{figure}[ht]
\centering
\begin{minipage}[b]{0.5\textwidth}\centering
\includegraphics[width=0.48\textwidth]{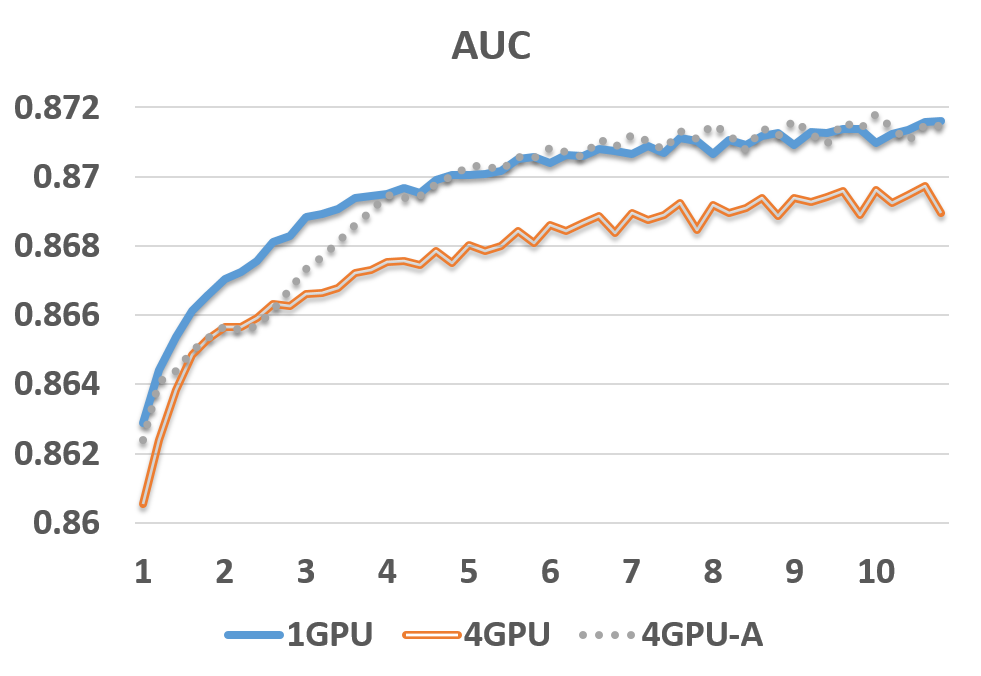}
\includegraphics[width=0.48\textwidth]{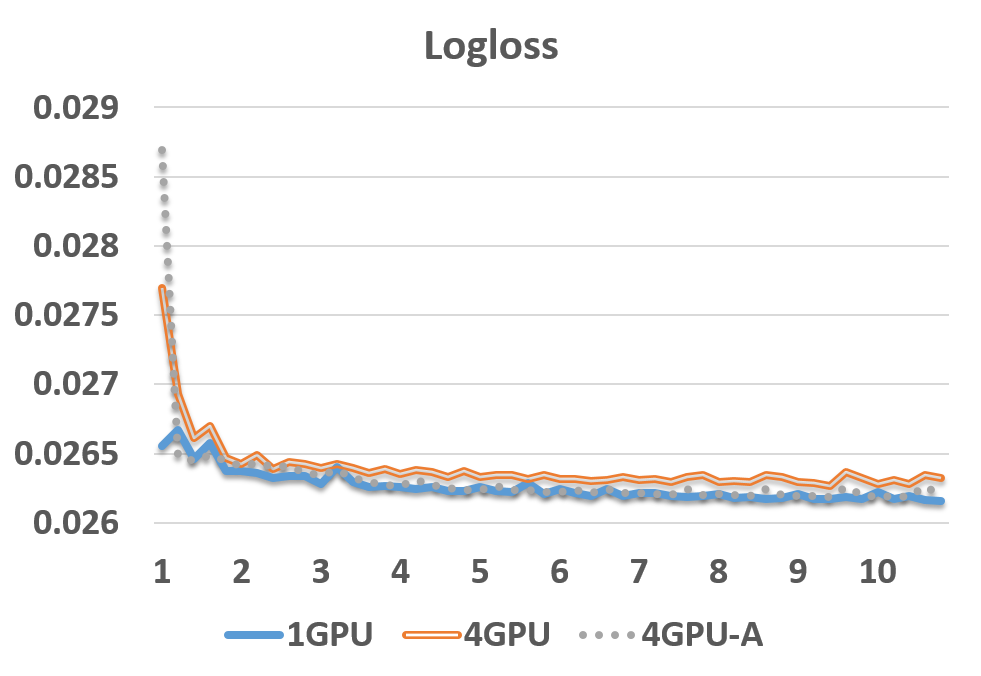}
\end{minipage}
\caption{\small{Performance comparison of multi-GPU data parallelism for DeepFM-D. Note: 1GPU means single GPU; 4GPU means 4 GPUs; 4GPU-A means 4 GPUs with adjusted $lr$.}}\label{fig:multi-cards}
\end{figure}
We evaluate the effectiveness of multi-GPU data parallelism for DeepFM-D model on Company$\ast$ data set, and the test curves of AUC and Logloss related with different settings are shown in Figure~\ref{fig:multi-cards}. Specifically, the batch size $bs$ of different settings are same, the learning rate $lr$ of 1GPU, 4GPU and 4GPU-A are 0.0001, 0.0001 and 0.0001$\times \sqrt{4}$, respectively. Compared with 1GPU, the test curves of 4GPU indicate that the training process of 4GPU is slower. It is because the number of updates of 4GPU is only a quarter of 1GPU when adopting same $bs$. As a result, it converges slower than 1GPU if we set same $lr$. In fact, the variance of the gradient in a mini-batch can be denoted as following,


\begin{displaymath}
\centering
\label{eq:variance}
Var(g)=Var(\frac{1}{bs}\sum_{\rm i=1}^{bs}g(x_{\rm i},y_{\rm i}))=\frac{1}{bs}Var(g(x_1,y_1)),
\end{displaymath}
where $g(x_{\rm i},y_{\rm i})$ is the gradient of a randomly selected instance. The reason for the second equal sign is that the variance of the gradient related to the randomly selected instances is equal to each other~\cite{DBLP:journals/corr/KeskarMNST16}. So the variance of the gradient decreases $bs$ times when we increase the batch size by $bs$ times. In other word, the gradient becomes more accurate. Then we can increase $lr$ to accelerate the training process. Add $lr$ into the equation of the gradient's variance:
 \begin{displaymath}
\centering
\label{eq:lr-variance}
\frac{1}{bs}Var(\sqrt{bs}\times lr \times g(x_1,y_1))= Var(lr \times g(x_1,y_1)).
\end{displaymath}
Therefore, when using 4 GPU cards, we can increase the value of $lr$ by $\sqrt{4}$ times. As a result, the learning curve of 4GPU-A in Figure~\ref{fig:multi-cards} is similar as that of 1GPU.

\begin{figure}[ht]
\centering
\includegraphics[width=0.48\textwidth]{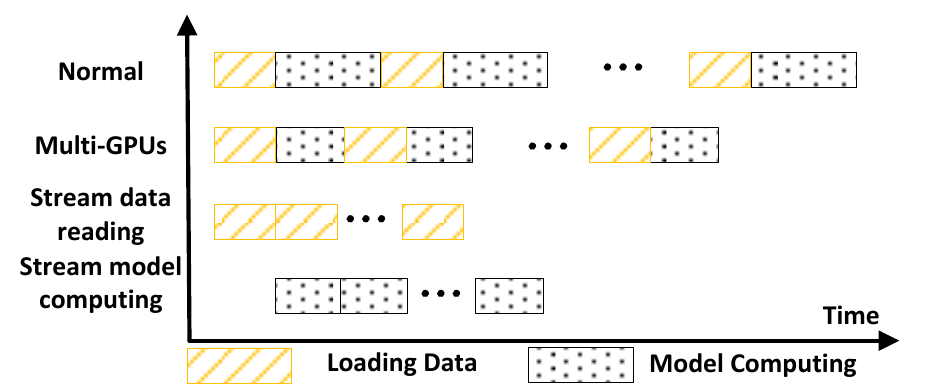}
\caption{The comparison of different strategies for loading data.}\label{fig:Asynchronous}
\end{figure}

\item \textbf{Asynchronous Data Reading}: Training a neural network is usually in a mini-batch style, in which a mini-batch of data records (e.g., several thousand) are read and fed into the neural network, and then model parameters are updated in a forward-backward way. There are two possible ways to handle data reading and model updating: sequential and parallel, as shown in Figure~\ref{fig:Asynchronous}. In the sequential approach, data reading and model updating are processed interleaved and sequentially, i.e., model updating starts when the current mini-batch of data is read and fed, and the next mini-batch of data will not be read until the model finishes updating with the current mini-batch of data. Obviously, this is not an efficient way. We propose a parallel manner to handle data reading and model updating, namely, \emph{asynchronous data reading}. A thread is created to read the data records regardless of the model updating, so that the model parameters keep updating without interrupted by reading data.

\item \textbf{Efficiency Validation Experiment}:

As shown in Table~\ref{table:valiadtion}, we record the speed up rate\footnote{In this paper, we define the speed up rate of strategy A over strategy B to be the processing time of strategy B divided by the process time of strategy A.} of the 4GPU over 1GPU data parallelism, asynchronous (Asyn) over synchronous (Syn) data reading of DeepFM models on Company$\ast$ and Criteo data sets. We omit the validation result on Criteo-Sequential data set, since Criteo-Sequential and Criteo-Random come from the same original data set with different splitting strategies.
\begin{table}[ht]
\centering
\small
\caption{\small{Speed up rate for DeepFM-D model.}}\label{table:valiadtion}
\begin{tabular}{ccc}
\toprule
                   & Company$\ast$ & Criteo-Random \\
\midrule
 4GPU over 1GPU    & 2.25 X          & 2.15 X   \\
 Asyn over Syn     & 1.12 X          & 1.19 X  \\
\bottomrule
\end{tabular}
\end{table}

\end{itemize}
\subsection{Relationship with Other Neural Networks}\label{section:App:rela}

Inspired by the enormous success of deep learning in various applications, several deep models for CTR prediction are developed recently. This section compares the proposed DeepFM-D and DeepFM-P models with existing deep models for CTR prediction.

\begin{figure*}[ht]
\centering
\includegraphics[width=1\textwidth]{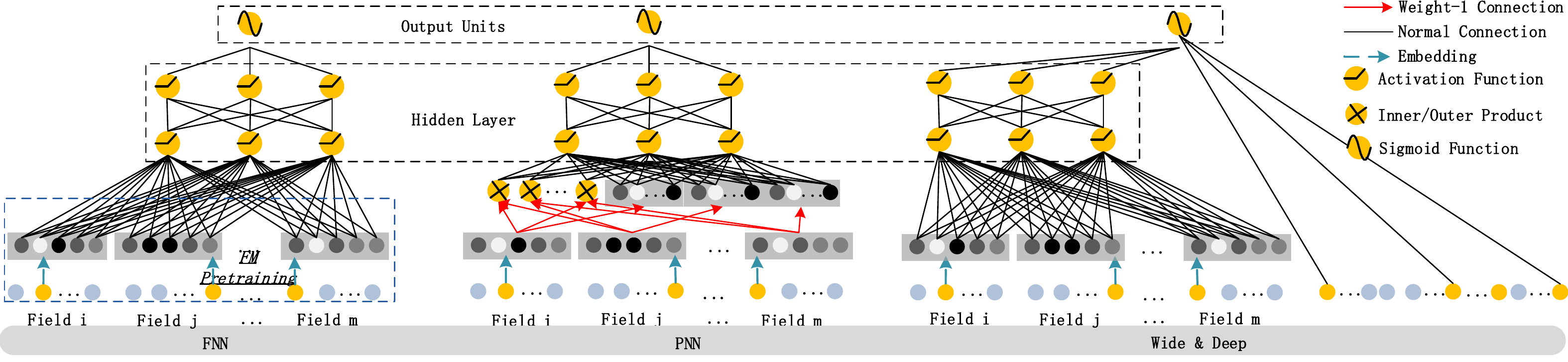}
\caption{The architectures of existing deep models for CTR prediction: FNN, PNN, Wide \& Deep Model.}\label{fig:othermodel}
\end{figure*}

\subsubsection{FNN}
As Figure~\ref{fig:othermodel} (left) shows, FNN is an FM-initialized feed-forward neural network~\cite{fnn}. The FM pre-training strategy results in two limitations: 1) the embedding parameters might be over affected by FM; 2) the efficiency is reduced by the overhead introduced by the pre-training stage. In addition, FNN captures only high-order feature interactions. In contrast, DeepFM-D and DeepFM-P models need no pre-training and learn both high- and low-order feature interactions in an end-to-end manner.

There is a detailed issue about embedding vectors among different models that needs to be metioned. As shown in Figure 1 of~\cite{fnn}, each embedding vector of a field includes both latent vector $V$ of this field and an additional neuron representing the weights of the feature values in this field. In other words, if the dimension of latent vectors in FM is $k$, then the latent vectors are of size $k+1$ in FNN. In our experiments, in order to keep the same representation ability, PNN and Wide \& Deep use the same size of embedding vectors as FNN. On the other side, DeepFM models have an FM component to model the weights of individual feature values, therefore there is no need to include an additional neuron in the embedding vector as FNN does. Due to this reason, the size of embedding vectors in DeepFM models is smaller than that in FNN by one.

\subsubsection{PNN}
For the purpose of capturing high-order feature interactions, PNN imposes a product layer between the embedding layer and the first hidden layer~\cite{pnn}. According to different types of product operations, there are three variants: IPNN, OPNN, and PNN$\ast$, where IPNN is based on inner product of vectors, OPNN is based on outer product, and PNN$\ast$ is based on both inner and outer products. To make the computation more efficient, the authors proposed the approximated computations of both inner and outer products: 1) the inner product is approximately computed by eliminating some neurons; 2) the outer product is approximately computed by compressing $m$ $k$-dimensional feature vectors to one $k$-dimensional vector. However, we find that the outer product is less reliable than the inner product, since the approximated computation of outer product loses much information that makes the result unstable. Although inner product is more reliable, it still suffers from high computational complexity, because the output of the product layer is connected to all neurons of the first hidden layer. Like FNN, all PNNs ignore low-order feature interactions.


\subsubsection{Wide \& Deep}
Wide \& Deep (Figure~\ref{fig:othermodel} (right)) is proposed by Google to model low- and high-order feature interactions simultaneously. As shown in~\cite{wide-n-deep}, there is a need for expertise feature engineering on the input to the ``wide" part (for instance, cross-product of users' install apps and impression apps in app recommendation). In contrast, DeepFM-D and DeepFM-P need no such expertise knowledge to handle the input by learning directly from the input raw features.

A straightforward extension to this model is replacing LR by FM (we also evaluate this extension in Section~\ref{section:exp}). This extension is similar to DeepFM-D, but DeepFM framework shares the feature embedding between the FM component and deep component. The sharing strategy of feature embedding influences (in back-propagate manner) the feature representation by both low- and high-order feature interactions, which models the representation more precisely.

\subsubsection{Summarizations}
To summarize, the relationship between DeepFM framework and the other deep models in four aspects is presented in Table~\ref{table:Modelcompare}. DeepFM is the only framework that requires no pre-training and no feature engineering, and captures both low- and high-order feature interactions.

\begin{table}
\centering
\scriptsize
\caption{{Comparison of deep models for CTR prediction.}}\label{table:Modelcompare}
\begin{tabular}{ccccc}
\toprule
 &  No& High-order  & Low-order  & No Feature \\
 &Pre-training     & Features     & Features    & Engineering \\
 \midrule
FNN & $\times$ & $\surd$  & $\times$ & $\surd$ \\
PNN & ${\surd}$ & ${\surd}$  & ${\times}$ & ${\surd}$ \\
Wide \& Deep & ${\surd}$   & ${\surd}$  & ${\surd}$  & ${\times}$\\
DeepFM-D (P) & ${\surd}$ & ${\surd}$  & ${\surd}$  & ${\surd}$ \\
\bottomrule
\end{tabular}
\end{table}

\section{Experiments}\label{section:exp}

In this section, we conduct both offline and online experiments to evaluate our proposed DeepFM framework.

In the offline experiments, we compare two instances of our proposed DeepFM framework (namely, DeepFM-D and DeepFM-P) with the other state-of-the-art models empirically. The evaluation result indicates that DeepFM-D and DeepFM-P are more effective than any other state-of-the-art model. The efficiency tests of DeepFM-D, DeepFM-P and the other baseline models are also performed.

In the online experiments, we conduct a consecutive seven days' A/B test to evaluate the performance of DeepFM framework. In these DeepFM models, DeepFM-D has a relative better efficiency and performance. Therefore, we adopt DeepFM-D as our model to compare with a well-engineered LR model, which is one of the most popular CTR prediction model in industry. In addition, to understand the result of A/B test better, we compare the recommendation lists generated by LR and DeepFM-D through the online simulation experiment.

\subsection{Setup of Offline Experiments}\label{sec:exp:set}

\subsubsection{Data sets}\label{sec:exp:set:data}

We evaluate the effectiveness and efficiency of our DeepFM-D and DeepFM-P models on the following three data sets.

\begin{itemize}[leftmargin=*]
\item \textbf{Criteo Data set}: Criteo data set\footnote{http://labs.criteo.com/downloads/2014-kaggle-display-advertising-challenge-dataset/} includes 45 million users' click records. There are 13 continuous features and 26 categorical ones. We split the data set in two different ways: randomly and sequentially, resulting in \emph{Criteo-Random} and \emph{Criteo-Sequential}. To get Criteo-Random data set, the original Criteo data set is randomly split into two parts as: 9/10 is for training, while the rest 1/10 is for testing. To get Criteo-Sequential data set, the original data set is split sequentially as: the first 6/7 is for training, while the rest 1/7 is for testing, as the original  data set consists of data instances of 7 consecutive days. In Criteo-Sequential data set, information is not leaked, however data is significantly biased between training set and test set, as training set contains only six days' records instead of one week's records. On the contrary, in Criteo-Random data set, information may be leaked but it is not significant biased between training set and test set.

\item \textbf{Company$\ast$ Data set}: Company$\ast$ data set is a commercial industrial data set. We collect 8 consecutive days of users' click records from the game center of the Company$\ast$ App Store: the first 7 days' records for training, and the next 1 day's records for testing. There are around 1 billion records in the whole collected dataset. In this dataset, there are app features (e.g., identification, category, and etc), user features (e.g., user's downloaded apps, and etc), and context features (e.g., operation time, and etc).
\end{itemize}
\subsubsection{Evaluation Metrics}\label{sec:exp:set:metric}

We use two evaluation metrics in our experiments: \textbf{AUC} (\textbf{A}rea \textbf{U}nder ROC \textbf{C}urve) and \textbf{Logloss} (\textbf{Log}istic \textbf{loss}).

AUC and Logloss are two of the most commonly used evaluation metrics for binary-class classification problem. For such machine learning models of binary-class classification, the prediction is a probability value that the given data record belongs to a certain class. AUC and Logloss are more suitable than precision and recall, it is because that when computing precision and recall, a user-defined threshold is needed to convert a probability value to a class label and the choice of the threshold value affects the accuracy and recall significantly. However, AUC and Logloss avoid such user-defined threshold values.


AUC is equal to the probability that a classifier will rank a randomly chosen positive instance higher than a randomly chosen negative one (assuming positive ranks higher than negative)~\cite{auc}. Logloss (or Cross Entropy) measures the distance between two distributions, one of which is predicted by the model while the other is given by the labels of the data instances. Note that Logloss is the objective function of our proposed model at the same time. The formula of Logloss is presented in Equation~\ref{eq:logloss}.


\subsubsection{Model Comparison}\label{sec:exp:set:comparemodel}

In our experiment, we compare the performance of 12 models, which are divided into four categories: \emph{Wide models}, \emph{Deep models}, \emph{Wide \& Deep models} and \emph{DeepFM models}.
\begin{itemize}[leftmargin=*]
\item Wide models: \textbf{LR}, \textbf{FM}.
\item Deep models: \textbf{DNN}, \textbf{FNN}, \textbf{PNN}. There are three variants of PNN, namely \textbf{IPNN}, \textbf{OPNN} and \textbf{PNN$\ast$}.
\item Wide \& Deep models: The original Wide \& Deep model is discussed in Section~\ref{section:App:rela}. For the purpose of eliminating feature engineering effort, we adapt the original Wide \& Deep model by replacing LR by FM as the wide part. In order to distinguish these two variants of Wide \& Deep, we name them \textbf{LR \& DNN} and \textbf{FM \& DNN}, respectively.\footnote{We do not use the Wide \& Deep API released by Google, as the efficiency of that implementation is very low. We implement Wide \& Deep by ourselves by simplifying it with shared optimizer for both deep and wide part.}
\item DeepFM models: \textbf{DeepFM-D} and \textbf{DeepFM-P}. Our DeepFM-P also has three variants, denoted as \textbf{DeepFM-IP}, \textbf{DeepFM-OP} and \textbf{DeepFM-$\ast$P}, of which the deep components are the three variants of PNN accordingly.
\end{itemize}

\subsubsection{Parameter Settings}\label{sec:exp:set:hyper}

To achieve the best performance for each individual model on Criteo-Random, Criteo-Sequential and Company$\ast$ data sets, we conducted carefully parameter study on all the data sets. Due to the space limit, we only discuss the parameter study on Company$\ast$ data set, which is presented in Section~\ref{sec:exp:hyper}. The hyper-parameters of compared deep models on Criteo-Sequential and Criteo-Random data sets are stated in Table~\ref{table:paraset}, where the activation function, dropout and structure of hidden layers are given. The optimizer for LR  and others are FTRL~\cite{FTRL} and Adam~\cite{DL_book}, respectively. The embedding dimensions of FM and DeepFM models are 10, and others are 11 (discussed in Section~\ref{section:App:rela}). Note that hyper-parameters of baseline models on Criteo-Random data set follow the setting in~\cite{pnn}, and we keep the deep components of DeepFM models with the same setting to validate the superiority of our models.
\begin{table}
\centering
\scriptsize
\caption{\small{Hyper-parameters of deep models on Criteo.}}\label{table:paraset}
\begin{tabular}{ccc}
\toprule
Model & Criteo-Random & Criteo-Sequential  \\
\midrule
DNN &relu;0.5;400-400-400.&relu;0.8;800-800-800-800-800. \\
FNN & relu;0.5;400-400-400.&relu;0.9;1100-1000-900-800-700-600-500. \\
IPNN &tanh;0.5;400-400-400. &tanh;0.8;800-800-800. \\
OPNN & relu;0.5;400-400-400.&relu;0.9;800-800-800-800-800. \\
PNN$\ast$& relu;0.5;400-400-400.&relu;0.8;800-800-800-800-800. \\
LR \& DNN & relu;0.5;400-400-400.&relu;1.0;1100-1000-900-800-700-600-500. \\
FM \& DNN &relu;0.5;400-400-400. &tanh;0.7;1000-900-800-700-600. \\
DeepFM-D & relu;0.5;400-400-400.  &relu;0.9;800-800-800-800-800-800-800.\\
DeepFM-IP &relu;0.5;400-400-400. &relu;0.8;800-800-800.\\
DeepFM-OP &relu;0.5;400-400-400. &relu;0.9;800-800-800-800-800.\\
DeepFM-$\ast$p &relu;0.5;400-400-400. &relu;0.8;800-800-800-800-800. \\
\bottomrule
\end{tabular}
\end{table}

\subsection{Performance of Offline Evaluation}\label{sec:exp:perfor}

In this section, we evaluate the efficiency and effectiveness of the models listed in Section~\ref{sec:exp:set:comparemodel} on the three data sets.

\subsubsection{Efficiency Comparison}\label{sec:exp:perfor:time}

\begin{figure}[ht]
\centering
\begin{minipage}[b]{0.5\textwidth}\centering
\includegraphics[width=0.75\textwidth]{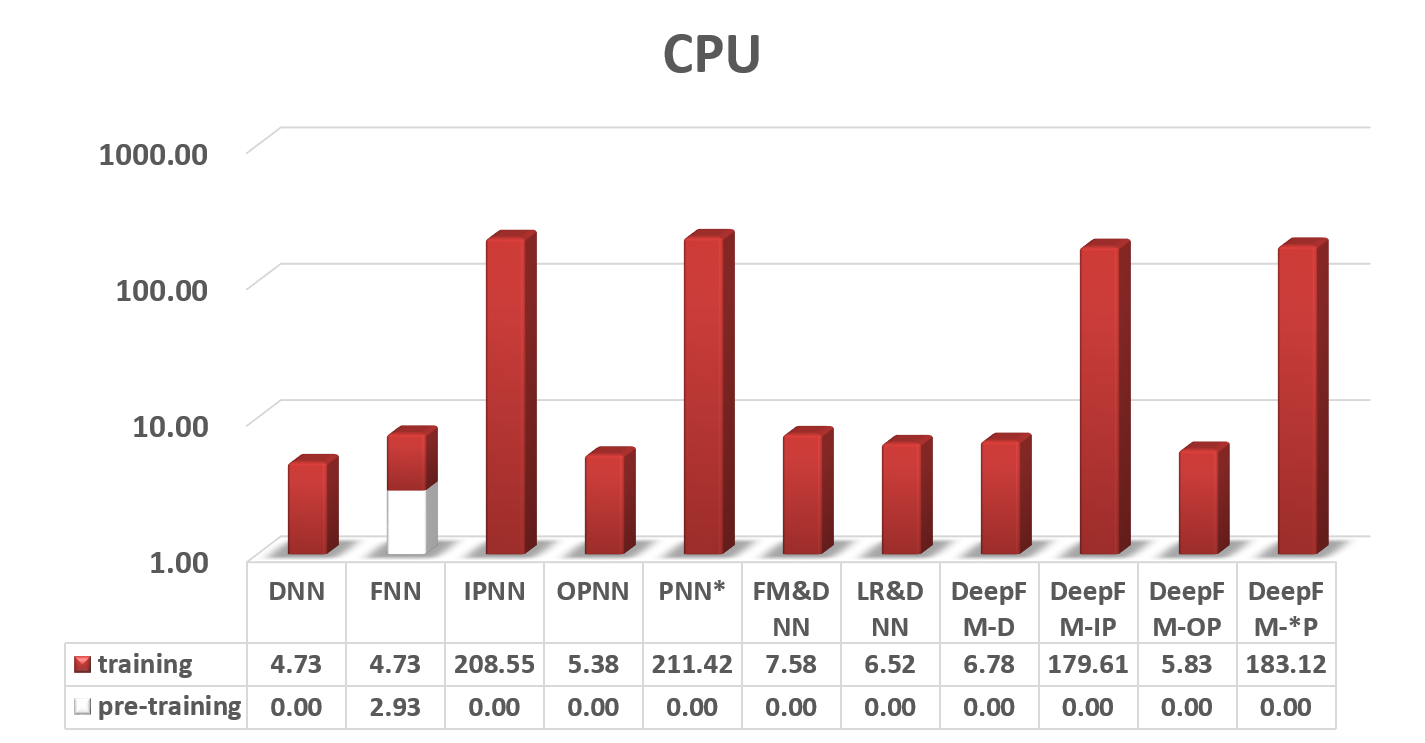}
\includegraphics[width=0.75\textwidth]{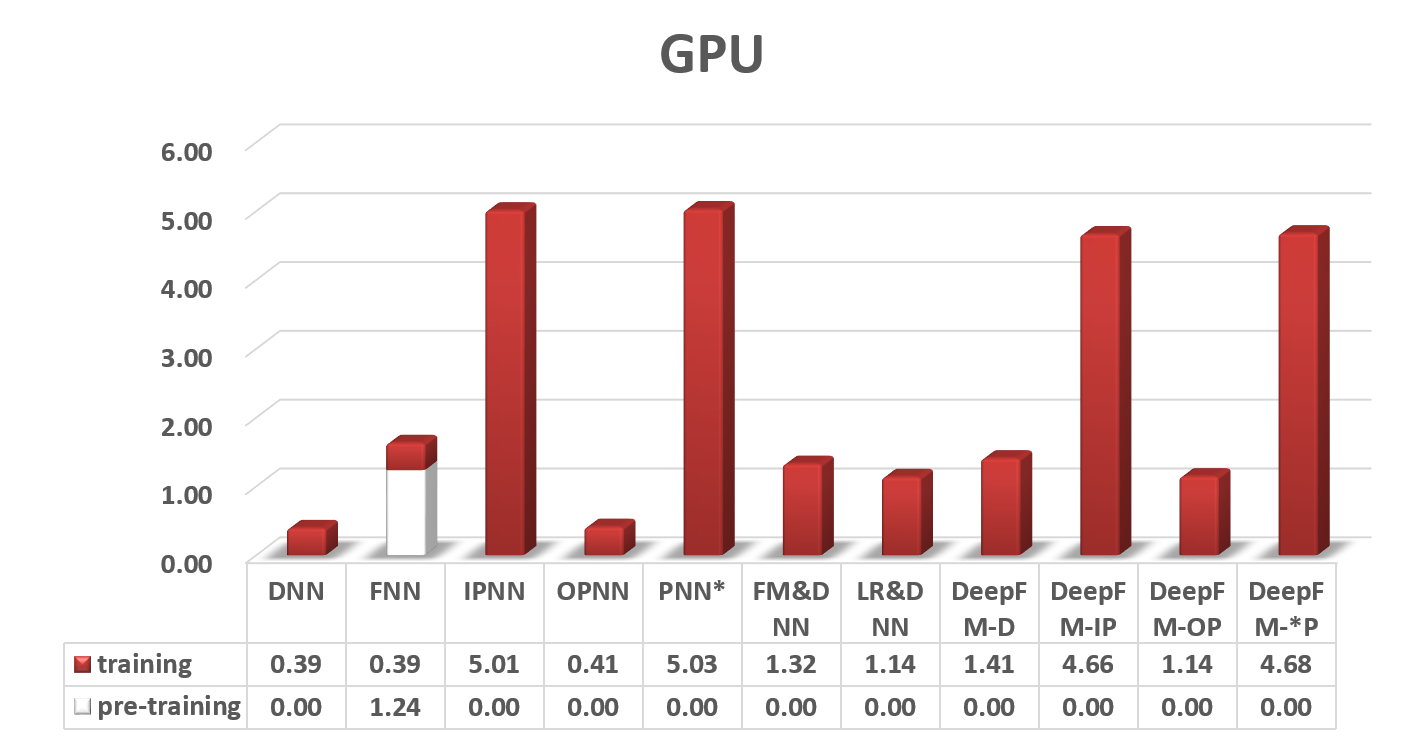}
\end{minipage}
\caption{Running time comparison between CPU and GPU.}\label{fig:time}
\end{figure}

The efficiency of deep learning models is important to real-world applications. We compare the efficiency of different models on Company$\ast$ data set by the following formula: $\frac{\rm |training\ time\ of\ deep\ CTR\ model|}{\rm |training\ time\ of\ LR|}$, which is the normalized running time by LR model. The results are shown in Figure~\ref{fig:time}, including the tests on CPU and GPU, where we have the following observations:
\begin{itemize}[leftmargin=*]
\item Pre-training of FNN makes it less efficient, especially on GPU, since the pre-training by FM model is not suitable for accelerating by GPU.
\item IPNN, PNN$\ast$, DeepFM-IP and DeepFM-$\ast$P are the least efficient models, on both CPU and GPU. Although their speed up on GPU is higher than the other models, they are still computationally expensive because of the inefficient inner product operations.
\item DNN and OPNN are the most efficient models, on both CPU and GPU. Testing on GPU shows a much more obvious gap between these two models and other models.
\item FNN, FM \& DNN, LR \& DNN, DeepFM-D and DeepFM-OP have similar efficiency on both CPU and GPU.
\end{itemize}

In real industry applications, we are equipped by powerful servers with GPUs. The efficiency of DeepFM-D and DeepFM-OP is acceptable for us, since they are only 41\% and 14\% slower than LR model on GPU.

\subsubsection{Effectiveness Comparison}

\begin{table*}[ht]
\centering
\small
\caption{\small{Performance (in terms of AUC and Logloss) of all the compared models.}}\label{table:performance}
\begin{tabular}{ccccccc}
\toprule
\multirow{2}{*}{} &\multicolumn{3}{c}{AUC} & \multicolumn{3}{c}{LogLoss} \\
 & Company$\ast$ & Criteo-Random & Criteo-Sequential & Company$\ast$ & Criteo-Random & Criteo-Sequential \\
 \midrule
LR      & 0.8641    & 0.7804 &0.7777 & 0.02648& 0.46782  &0.4794 \\
FM      & 0.8679    & 0.7894 &0.7843 & 0.02632&  0.46059  & 0.4739 \\
DNN     & 0.8650     &0.7860   &0.7953   & 0.02643    &0.4697      &0.4580  \\
FNN     & 0.8684    & 0.7959 &0.8038 & 0.02628&  0.46350 & 0.4507\\
IPNN    & 0.8662    & 0.7971 &0.7995 & 0.02639& 0.45347 & 0.4543\\
OPNN    & 0.8657    & 0.7981 &0.8002 & 0.02640& 0.45293  & 0.4536\\
PNN$\ast$&0.8663    & 0.7983 &0.8005 & 0.02638& 0.453305 & 0.4533\\
LR \& DNN & 0.8671  & 0.7858 &0.7973 & 0.02635& 0.46596  & 0.4565 \\
FM \& DNN & 0.8658  & 0.7980 &0.7985 & 0.02639& 0.45343  & 0.4551 \\
DeepFM-D  & 0.8715  &0.8016  &\textbf{0.8048} &0.02619  & 0.44985  &\textbf{0.4497}\\
DeepFM-IP & \textbf{0.8720}& \textbf{0.8019}   &0.8019   & \textbf{0.02616}& \textbf{0.4496} & 0.4525 \\
DeepFM-OP & 0.8713& 0.8008&   0.8020  & 0.02619&  0.4510  &0.4524 \\
DeepFM-$\ast$P & 0.8716& 0.7995  &0.8015  & 0.02619&  0.4515 &0.4530\\
\bottomrule
\end{tabular}
\end{table*}

The performance (in terms of AUC and Logloss) of the compared models on Criteo-Random data set, Criteo-Sequential data set and Company$\ast$ data set is presented in Table~\ref{table:performance} (the values in the table are averaged by 5 runs, and the variances of AUC and Logloss are in the order of $10^{-5}$). The following conclusions are observed:
\begin{itemize}[leftmargin=*]
\item Learning feature interactions improves the performance. LR, the only model that does not consider feature interactions, performs worse than the other models. As the best models, DeepFM models outperform LR by 0.91\%, 2.75\% and 3.48\% in terms of AUC (1.21\%, 3.89\% and 6.2\% in terms of Logloss) on Company$\ast$, Criteo-Random and Criteo-Sequential data sets respectively.
\item The performance of a DeepFM model is better than the model that keeps only the FM component or keeps only the Deep component. That is to say, the performance of DeepFM-D (DeepFM-IP, DeepFM-OP, DeepFM-$\ast$P, respectively) is better than both FM and DNN (IPNN, OPNN, PNN$\ast$, respectively). Table~\ref{table:improve-widedeep} presents performance improvement of the four DeepFM models over FM component and their deep components on the three data sets.
\item Learning high- and low-order feature interactions simultaneously and properly improves the performance. Among DeepFM models, DeepFM-D and DeepFM-IP perform the best. DeepFM-D and DeepFM-IP outperform the models that learn only low-order feature interactions (namely, LR and FM) or high-order feature interactions (namely, FNN, IPNN, OPNN, PNN$\ast$). Compared with the best baseline that learns high- or low-order feature interactions alone, DeepFM-D and DeepFM-IP achieve more than 0.41\%, 0.45\% and 0.12\% in terms of AUC (0.46\%, 0.82\% and 0.22\% in terms of Logloss) on Company$\ast$, Criteo-Random and Criteo-Sequential data sets.
\item Learning high- and low-order feature interactions simultaneously while sharing the same feature embedding for high- and low-order feature interactions learning improves the performance. DeepFM-D outperforms the models that learn high- and low-order feature interactions using separate feature embeddings (namely, LR \& DNN and FM \& DNN). DeepFM-D achieves more than 0.48\%, 0.44\% and 0.79\% in terms of AUC (0.58\%, 0.80\% and 1.2\% in terms of Logloss) on Company$\ast$, Criteo-Random and Criteo-Sequential data sets, respectively.
\end{itemize}

%

\begin{table*}[ht]
\centering
\footnotesize
\caption{{The imporvement of DeepFM models over its wide component (namely FM model) and deep component.}}\label{table:improve-widedeep}
\begin{tabular}{cccccccccc}
\toprule
&&\multicolumn{4}{c}{Wide}&\multicolumn{4}{c}{Deep} \\
 & & DeepFM-D & DeepFM-IP & DeepFM-OP & DeepFM-$\ast$P & DeepFM-D & DeepFM-IP & DeepFM-OP & DeepFM-$\ast$P\\
 \midrule
\multirow{2}{*}{Company$\ast$}&   AUC  &0.4\% & 0.47\% & 0.39\% &0.43\% &0.75\% & 0.67\% & 0.65\% &0.61\% \\
                              &   LogLoss  & 0.49\% &0.61\% &0.49\% &0.49\% & 0.91\% &0.87\% &0.80\% &0.72\%\\
\multirow{2}{*}{Criteo-Random} &  AUC     &1.54\% &1.58\% &1.44\% &1.28\% &1.98\% &0.60\% &0.34\% &0.15\% \\
&   LogLoss &2.33\% &2.39\% &2.08\% &1.97\% &4.22\% &0.85\% &0.43\% &0.40\%\\
\multirow{2}{*}{Criteo-Sequential}&  AUC & 2.61\% &2.24\% &2.26\% &2.19\%   & 1.19\% &0.30\% &0.22\% &0.12\% \\
&LogLoss &5.1\% &4.52\% &4.54\% &4.41\% &1.81\% &0.40\% &0.26\% &0.07\%\\
\bottomrule
\end{tabular}
\end{table*}


Overall, our proposed four DeepFM models perform better than the baseline models in all the cases. In particular, our proposed DeepFM-D\footnote{Although DeepFM-IP performs slightly better than DeepFM-D on Company$\ast$ dataset, we will still choose DeepFM-D in our real scenario to avoid the high time complexity of DeepFM-IP.} model beats the competitors by more than 0.36\% and 0.34\% in terms of AUC and Logloss on Company$\ast$ data set. In fact, a small improvement in offline AUC evaluation is likely to lead to a significant increase in online CTR. As reported in~\cite{wide-n-deep}, compared with LR, Wide \& Deep improves AUC by 0.275\% (offline) and the improvement of online CTR is 3.9\%.
Moreover, we also conduct t-test between our proposed DeepFM models and the baseline models on the three data sets. We find that the p-values are all less than $10^{-6}$, which indicates that our improvement over existing models is significant.


\subsection{Offline Hyper-Parameter Study}\label{sec:exp:hyper}

We study the impact of different hyper-parameters of different deep models, on Company$\ast$ dataset. The order is: 1) activation functions; 2) dropout rate; 3) number of neurons per layer; 4) number of hidden layers; 5) network shape. \emph{It can be clearly observed from the following sections that our proposed DeepFM models are significantly superior, compared with the baseline models, in all the studied cases.}

\subsubsection{Activation Function}\label{sec:exp:hyper:act}
\begin{figure}[ht]
\centering
\begin{minipage}[b]{0.5\textwidth}\centering
\includegraphics[width=0.85\textwidth]{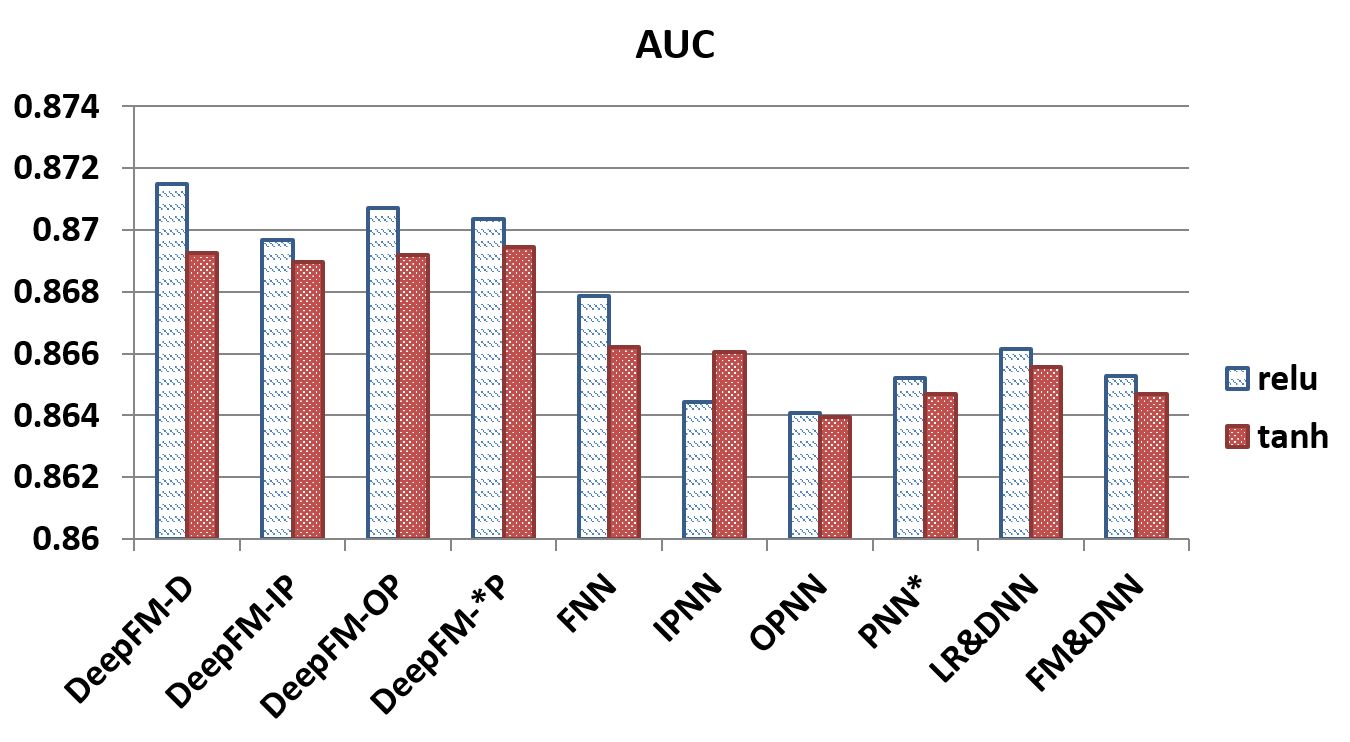}
\includegraphics[width=0.85\textwidth]{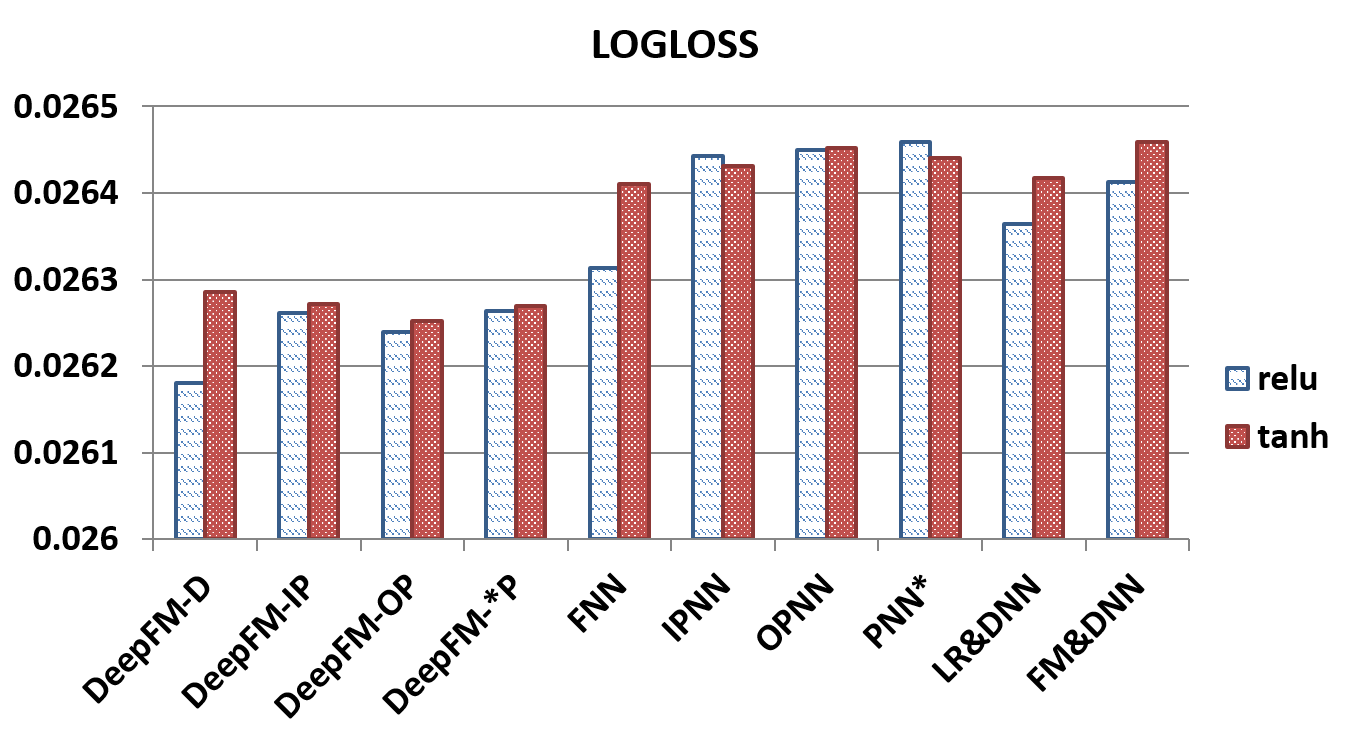}
\end{minipage}
\caption{\small{Performance comparison of activation functions in different models.}}\label{fig:act}
\end{figure}

According to \cite{pnn}, \emph{relu} and \emph{tanh} are more suitable for deep models than \emph{sigmoid}. The detailed discussion of different activation functions is presented in Section~\ref{subsection:learning}. In this paper, we compare the performance of deep models when applying \emph{relu} and \emph{tanh} as the activation function. As shown in Figure~\ref{fig:act}, relu is more appropriate than tanh for all the deep models, except for IPNN, due to the reason stated in Section~\ref{subsection:learning}.

\subsubsection{Dropout}\label{sec:exp:hyper:drop}
\begin{figure}[ht]
\centering
\begin{minipage}[b]{0.5\textwidth}\centering
\includegraphics[width=0.48\textwidth]{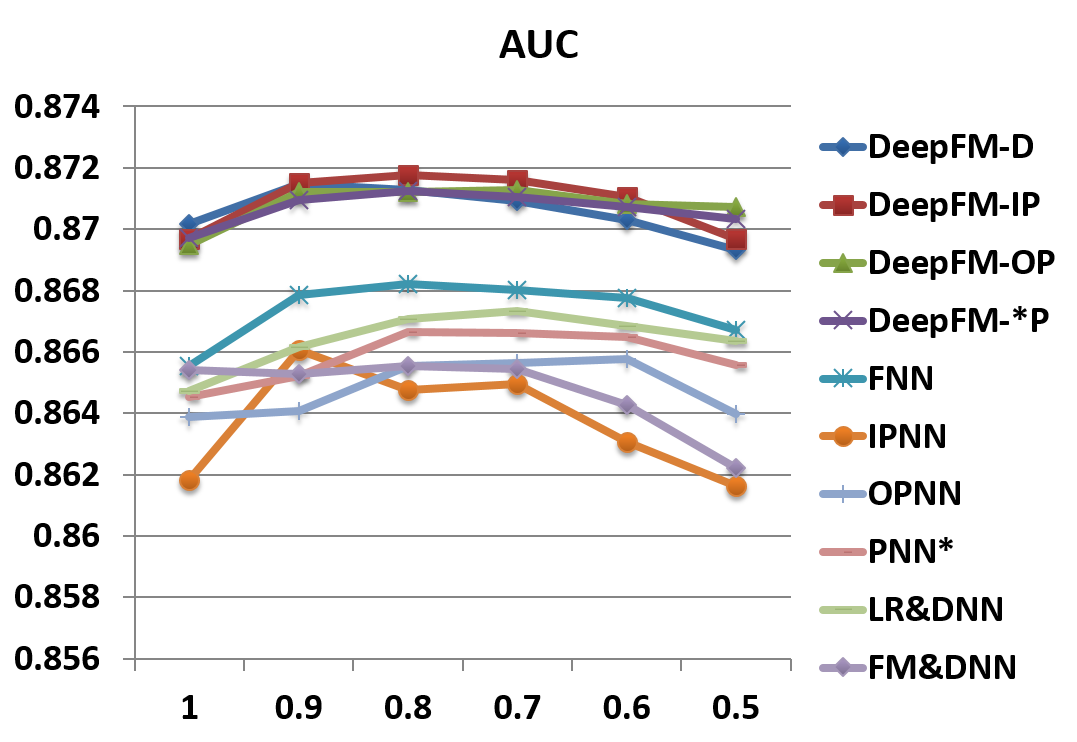}
\includegraphics[width=0.48\textwidth]{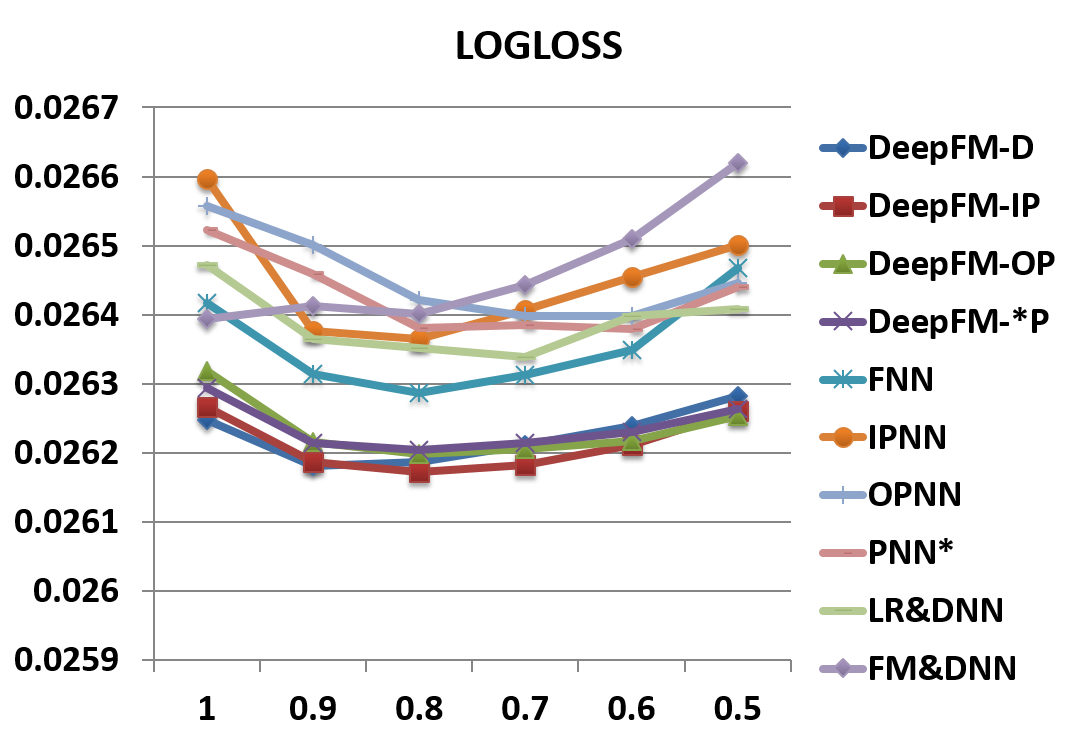}
\end{minipage}
\caption{{Performance comparison of dropout values in different models.}}\label{fig:drop}
\end{figure}

Dropout~\cite{dropout14} refers to the probability that a neuron is kept in the network. Dropout is a regularization technique to compromise the precision and the complexity of the neural network. We set the dropout to be 1.0, 0.9, 0.8, 0.7, 0.6, 0.5. As shown in Figure~\ref{fig:drop}, all the models are able to reach their own best performance when the dropout is properly set (from 0.6 to 0.9). The result shows that adding reasonable randomness to model can strengthen model's robustness and generalization.

\subsubsection{Number of Neurons per Layer}\label{sec:exp:hyper:neuron}
\begin{figure}[ht]
\centering
\begin{minipage}[b]{0.5\textwidth}\centering
\includegraphics[width=0.48\textwidth]{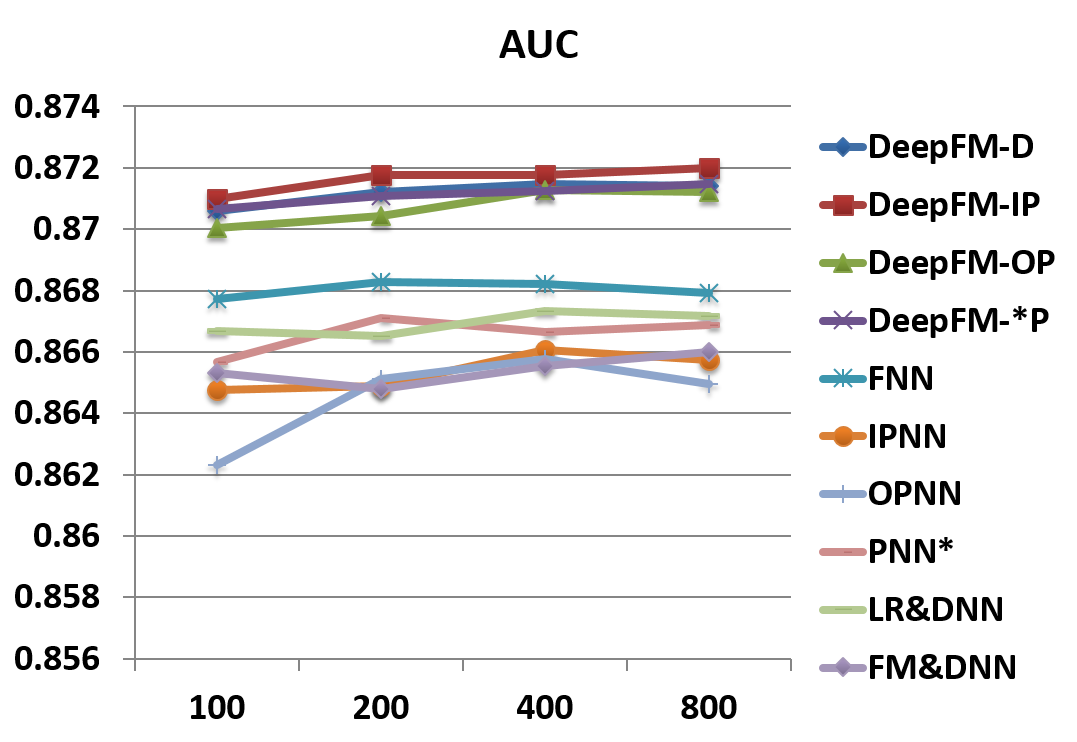}
\includegraphics[width=0.48\textwidth]{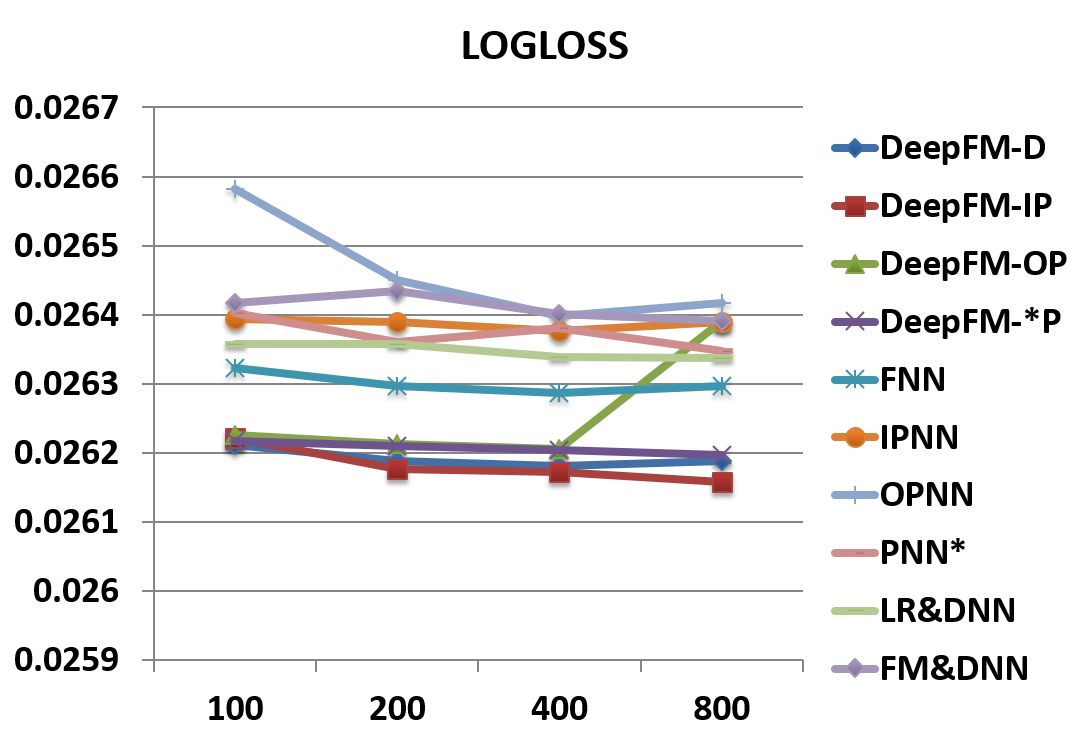}
\end{minipage}
\caption{{Performance comparison of number of neurons.}}\label{fig:neuron}
\end{figure}

When other factors remain the same, increasing the number of neurons per layer introduces complexity. When we study the effect of number of neurons per layer on the performance, we set the number of hidden layers to 3 and keep the number of neurons the same for each hidden layer. As observed from Figure~\ref{fig:neuron}, increasing the number of neurons does not always bring benefit. For instance, DeepFM-D performs stably when the number of neurons per layer is increased from 400 to 800; even worse, OPNN performs worse when we increase the number of neurons from 400 to 800. This is because an over-complicated model is easy to overfit. In our dataset, 200 or 400 neurons per layer is a good choice.

\subsubsection{Number of Hidden Layers}\label{sec:exp:hyper:layer}
\begin{figure}[ht]
\centering
\begin{minipage}[b]{0.5\textwidth}\centering
\includegraphics[width=0.48\textwidth]{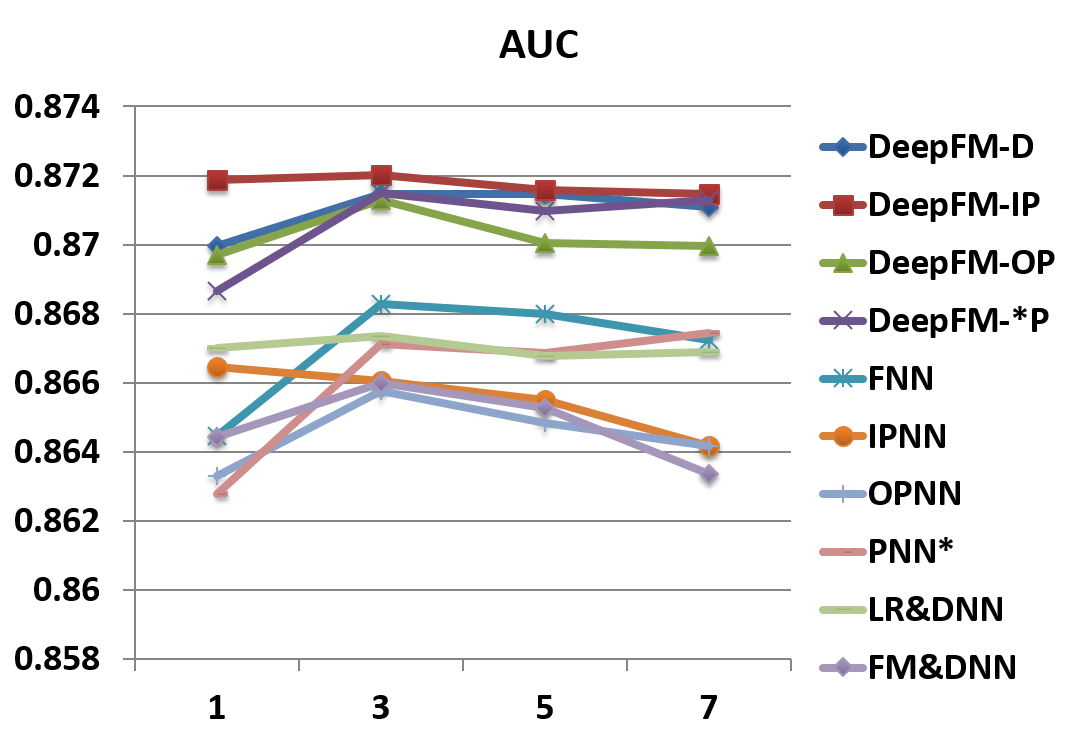}
\includegraphics[width=0.48\textwidth]{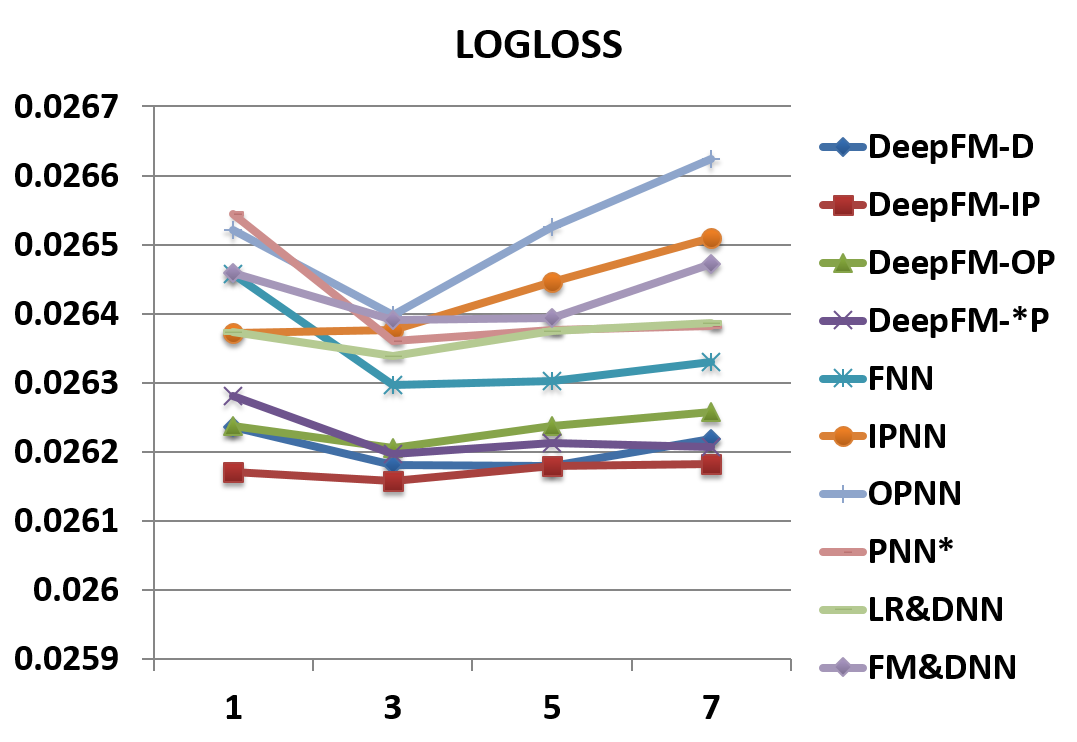}
\end{minipage}
\caption{{Performance comparison of number of layers in different models.}}\label{fig:layer}
\end{figure}

Varying the number of hidden layers, the number of neurons for each hidden layer is fixed. As presented in Figure~\ref{fig:layer}, increasing the number of hidden layers improves the model performance at the beginning, however, their performance is degraded if the number of hidden layers keeps increasing, because of overfitting.

\subsubsection{Network Shape}\label{sec:exp:hyper:shape}
\begin{figure}[ht]
\centering
\begin{minipage}[b]{0.5\textwidth}\centering
\includegraphics[width=0.48\textwidth]{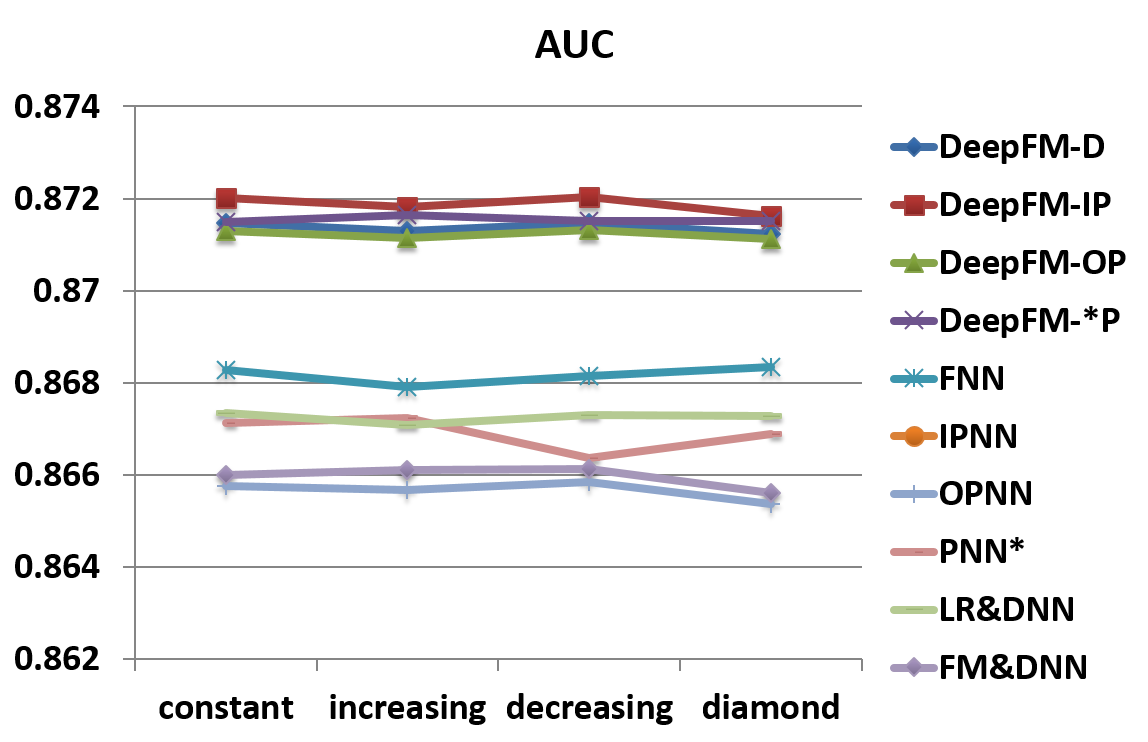}
\includegraphics[width=0.48\textwidth]{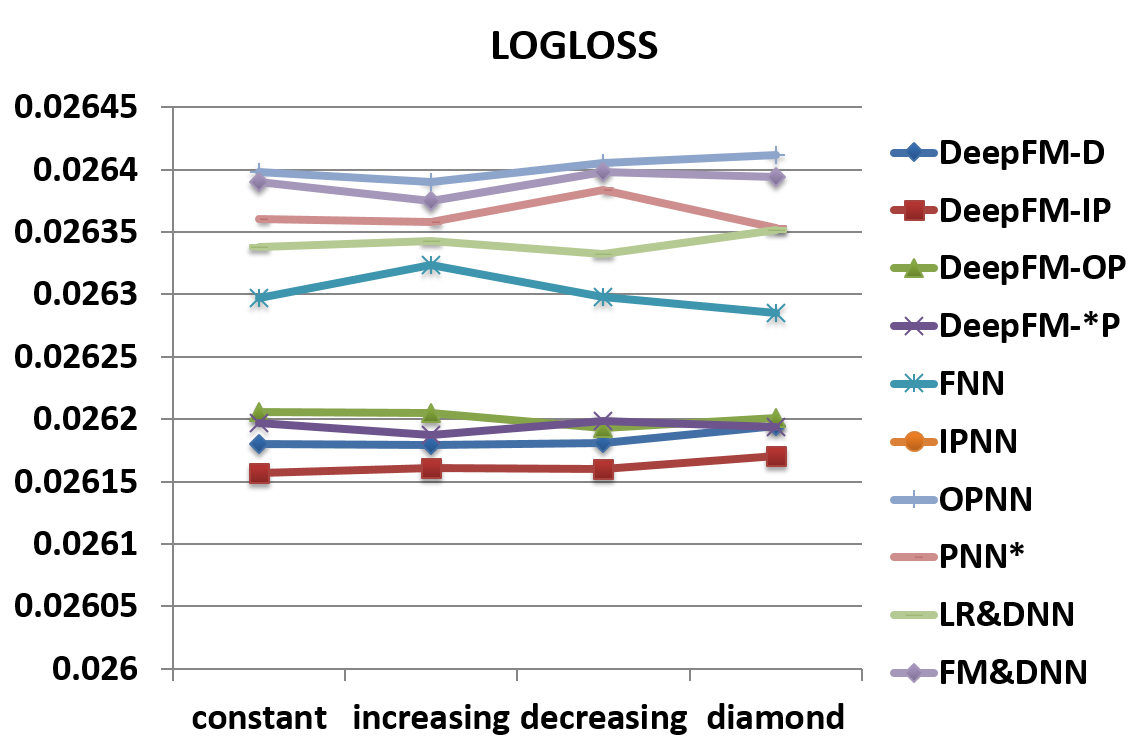}
\end{minipage}
\caption{{Performance comparison of network shape in different models.}}\label{fig:shape}
\end{figure}

We test four different network shapes: constant, increasing, decreasing, and diamond. When we change the network shape, we fix the number of hidden layers and the total number of neurons. For instance, when the number of hidden layers is 3 and the total number of neurons is 600, then four different shapes are: constant (200-200-200), increasing (100-200-300), decreasing (300-200-100), and diamond (150-300-150). As we can see from Figure~\ref{fig:shape}, the ``constant" network shape is empirically better than the other three options, which is consistent with previous studies~\cite{networkstructure09}.

\subsection{Online Experiments}\label{sec:exp:online}

According to the results of the offline experiments on Criteo-sequential, Criteo-random and Company$\ast$ datasets, the DeepFM models have shown their superior effectiveness over the other existing models in terms of AUC and Logloss. In order to verify the superior of DeepFM in a production environment, we implement and deploy DeepFM-D in the recommendation engine of Huawei App Market, which is one of the most popular Android App Markets in China.

Furthermore, we conduct two kinds of experiments to reveal the discriminative power of DeepFM-D compared to LR model, in CTR prediction task.

\begin{itemize}[leftmargin=1em]
\item  \textbf{A/B test:} Besides offline evaluation, it is more valuable to verify that whether DeepFM-D is able to perform its superiority as well in the real production environment of Huawei App Market. Therefore we conduct a consecutive seven days' A/B test, to test the performance of DeepFM-D, against a well-engineered LR model.

\item \textbf{Online simulation:} From a model perspective, compared with LR, DeepFM-D is able to capture high-order feature interactions, resulting in highly personalized recommendation. We aim to verify this statement with an online simulation, by analyzing the difference between recommendation results generated by DeepFM-D and LR.
\end{itemize}

In the following of this section, we give a brief description of the experiment settings in Section~\ref{sec:exp:online:setting}, then we present the results of online A/B test and online simulation in Section~\ref{sec:exp:online:abtest} and Section~\ref{sec:exp:online:simulation}, respectively.

\subsubsection{Setting}\label{sec:exp:online:setting}

In this section, we present the settings of A/B test and online simulation, including the experiment set up and evaluation metrics.

\begin{itemize}[leftmargin=1em]
\item \textbf{A/B test:} Considering the project launching schedule, we split the users into 2 groups, one group receives the recommendation by an LR model, which is one of the most popular CTR prediction model in industry; the other one gets the recommendation by DeepFM-D. The update frequencies of DeepFM and LR models are both on the daily basis. The A/B test is conducted on ``fun games" scenario in Huawei App Market. There are hundreds of millions of real users in Huawei App Market from whom consent has been acquired. After online A/B test in consecutive seven days, we collect the number of browsing and downloading records for both user groups in each day. We use CTR (Click Through Rate) and CVR (ConVersion Rate) as evaluation metrics:
 \begin{equation}
 CTR=\frac{\rm \sharp downloads}{\rm \sharp impressions},
 \end{equation}
 \begin{equation}
 CVR=\frac{\rm \sharp downloads}{\rm \sharp users},
 \end{equation}
 where $\rm \sharp downloads$ is the number of download records, $ \rm \sharp impressions$ is the number of browsing records, and $\rm \sharp users$ is the number of visited users.
\item \textbf{Online simulation:} Online simulation analyzes the properties of recommendation lists generated by LR and DeepFM-D. In order to study the difference between the recommendation results by LR and DeepFM-D in terms of personalization and diversity, we compare the cases for different types of users. Differentiated by users' downloading history, we generate $\rm t=6$ types of users, and each type includes $\rm n=100$ users. The user set is denoted as $U=\{U_{\rm ij}\}$, where $U_{\rm ij}$ represents the ${\rm j^{\rm th}}$ user of ${\rm i^{\rm th}}$ type and $U_{\rm i}$ represents all the users of type $\rm i$. A user of type $\rm i$ is generated by sampling several apps of type $\rm i$ as the user's downloading history. Then for user set $U$, we use the trained LR model (and DeepFM-D model) to generate recommendation list $R$, where $R=\{R_{\rm 11}, ..., R_{\rm tn}\}$ and $R_{\rm ij}$ is for user $U_{\rm ij}$. Based on the recommendation lists, we adopt \textbf{personalization@{\rm L}}~\cite{zhou2010solving}, \textbf{coverage@{\rm L}} and \textbf{popularity@{\rm L}}~\cite{SessionKNNJannachL17} as the evaluation metrics to investigate the differences between LR and DeepFM-D model.
    \begin{itemize}
    \item The \textbf{personalization@{\rm L}} considers the diversity of Top-${\rm L}$ places in different users' recommendation lists. We define the \emph{inter-list distance} between recommendation lists of user ${\rm a}$ and user ${\rm b}$ as
        \begin{equation}
         h_{\rm ab}=1-\frac{q_{\rm ab}({\rm L})}{\rm L},
        \end{equation}
        where $q_{\rm ab}(\rm L)$ is the common items in the Top-${\rm L}$ places of both lists. The \emph{inter-group distance} between recommendation lists of user group $U_{\rm i}$ and $U_{\rm j}$ is defined as the aggregated inter-list distances between the users across $U_{\rm i}$ and $U_{\rm j}$, as
        \begin{equation}
        h_{U_{\rm i}U_{\rm j}}=\frac{1}{|U_{\rm i}|\times|U_{\rm j}|}\sum_{{\rm a}\in U_{\rm i}}\sum_{{\rm b}\in U_{j}}h_{\rm ab},
        \end{equation}
        where $|U_{\rm i}|$ is cardinality of user group $U_{\rm i}$. Finally, the personalization of recommendation lists by a model is defined as aggregated inter-group distances between all pairs of user groups, as
        \begin{equation}
        h=\frac{2}{\rm t\times (t-1) }\sum_{\rm i=1}^{\rm t}\sum_{\rm j={\rm i+1}}^{\rm t}h_{U_{\rm i}U_{\rm j}}.
        \end{equation}
\item The \textbf{coverage@{\rm L}} considers the percentage of recommended apps in Top-${\rm L}$ places of all the users' recommendation lists over all the candidate apps.
        \begin{equation}
        coverage@{\rm L}=\frac{|{\bigcup_{\rm i\in[1,t],j\in[1,n]}{R_{\rm ij}({\rm L})}}|}{\rm |candidate~apps|},
        \end{equation}
        where ${R_{\rm ij}({\rm L})}$ is the recommended apps in Top-${\rm L}$ places in $R_{\rm ij}$.
\item In addition, we also measure the \textbf{popularity@{\rm L}}, which is defined as:
        \begin{equation}
        popularity@{\rm L}=\frac{1}{\rm t\times n \times {\rm L}}\sum_{\rm i\in[1,t],j\in[1,n]}{\sum_{{\rm k} \in R_{\rm ij}({\rm L})}\frac{D_{\rm k}}{D_{max}}},
        \end{equation}
      where $D_{\rm k}$ is the number of historical cumulative downloads of the recommended app $\rm k$ in $R_{\rm ij}({\rm L})$, $D_{max}$ is the number of historical cumulative downloads of the most downloaded app.
\end{itemize}
\end{itemize}

\subsubsection{Performance of Online Experiments}\label{sec:exp:online:abtest}

\begin{figure}[ht]
\centering
\includegraphics[width=0.35\textwidth]{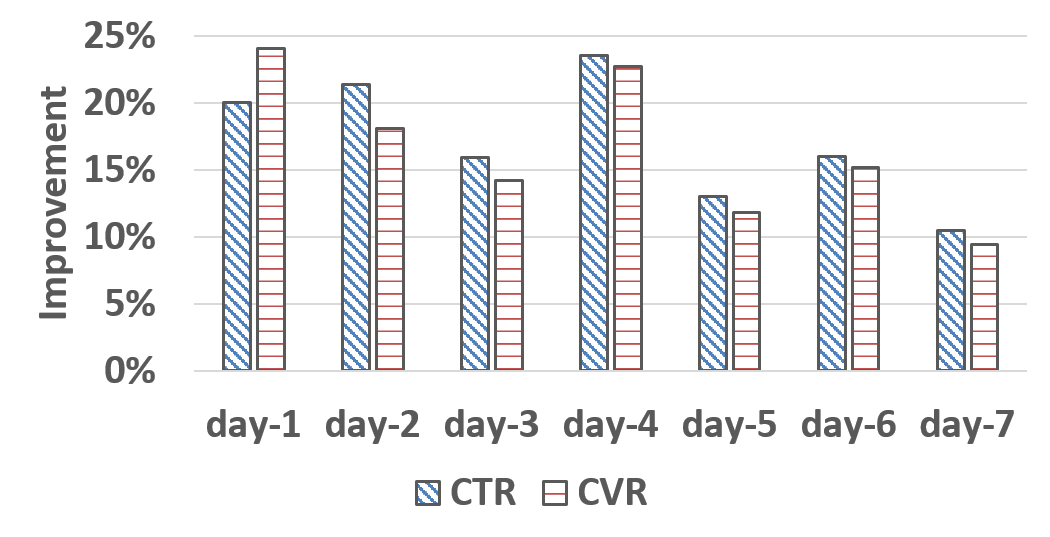}
\caption{The online CTR and CVR improvement of DeepFM-D over LR.}\label{fig:online-ctr}
\end{figure}

In this section, we present the results of online A/B test. Because of commercial concerns, we only report the improvements of DeepFM-D over LR in terms of CTR and CVR, as shown in Figure~\ref{fig:online-ctr}. The x-axis represents different days, and the y-axis is the improvement of DeepFM-D over LR. Note that the blue bar with slash line represents the improvement of CTR, while the red bar with horizonal line represents the improvement of CVR.

The histograms shows that the performance of DeepFM-D is consistently better than LR, through the whole A/B testing period. Specifically, the improvements of DeepFM-D over LR are at least 10\% in terms of CTR and CVR, except for the CVR on day-7 (which is still very close to 10\%). In addition, the highest improvement of CTR reaches about 24\% on the day-4 and the maximum of CVR improvement is about 25\% on day-1. The online A/B test results reveal that DeepFM-D leads to a higher CTR and CVR over LR in a recommendation engine of industry scale.

\subsubsection{The Property of Online Recommendations}\label{sec:exp:online:simulation}

\begin{figure}[ht]
\centering
\includegraphics[width=0.35\textwidth]{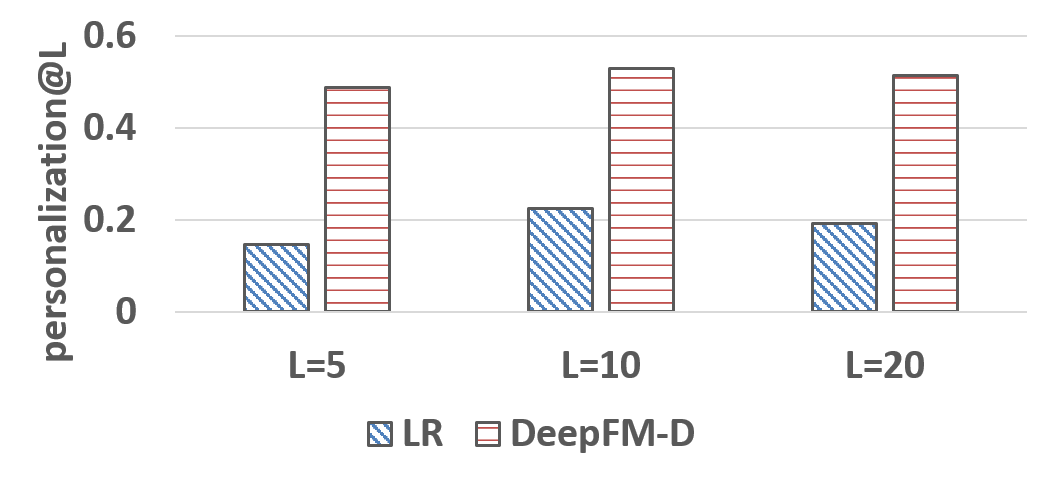}
\caption{The comparison of personalization of Top-L recommendation lists between DeepFM-D and LR.}\label{fig:online-diff}
\end{figure}

\begin{figure}[ht]
\centering
\includegraphics[width=0.35\textwidth]{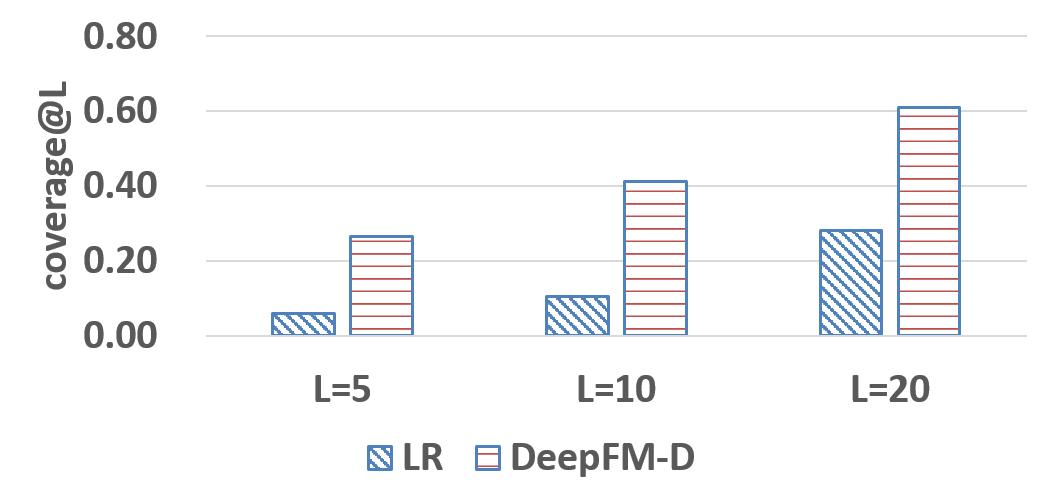}
\caption{The comparison of coverage of Top-L recommendation lists between DeepFM-D and LR.}\label{fig:online-coverage}
\end{figure}

\begin{figure}[ht]
\centering
\includegraphics[width=0.35\textwidth]{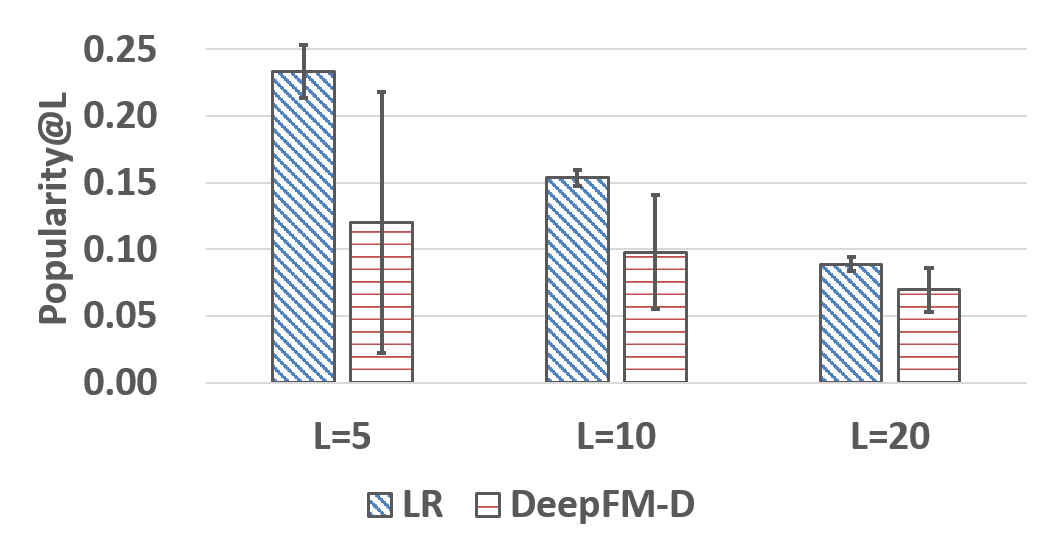}
\caption{The comparison of average degree and standard variance of Top-L recommendation lists between DeepFM-D and LR.}\label{fig:online-degree}
\end{figure}

To better understand the results of the online experiments, we conduct online simulation experiments and make a comparison of \textbf{personalization@{\rm L}}, \textbf{coverage@{\rm L}} and \textbf{popularity@{\rm L}} ($\rm L=\{5, 10, 20\}$) between the recommendation lists by LR and DeepFM-D. The results are presented in Figure~\ref{fig:online-diff}, Figure~\ref{fig:online-coverage} and Figure~\ref{fig:online-degree}, respectively. In these three figures, the blue bar with slash line and the red bar with horizonal line represent the measurement of recommendation lists by LR and DeepFM-D, respectively.

As shown in Figure~\ref{fig:online-diff} and Figure~\ref{fig:online-coverage}, the personalization@{\rm L} and coverage@{\rm L} of recommendation lists generated by DeepFM-D are much larger than that of LR. The personalization@{\rm L} is the aggregated inter-group distance between recommendation lists of different user groups, therefore a low personalization@{\rm L} means Top-${\rm L}$ places in the recommendation lists of different users are similar. The coverage@{\rm L} has similar semantics as the personalization@{\rm L}. When coverage@{\rm L} is low, the Top-${\rm L}$ places in the recommendation lists concentrate in a small range of apps. The results of personalization@{\rm L} and coverage@{\rm L} demonstrate that the Top-${\rm L}$ places in recommendation lists of DeepFM-D are more diverse than that of LR.

Figure~\ref{fig:online-degree} presents the comparison of popularity@{\rm L} of recommendation lists generated by LR and DeepFM-D. There are two indices in this experiment, the average and the variance of the historical cumulative downloads of the apps contained in the Top-${\rm L}$ places in recommendation lists. Specifically, the black line on the top of each bar is the variance of the 600 recommendation lists (100 users per type $\times$ 6 types). Compared with DeepFM-D, LR generates the recommendation lists with higher average popularity and lower variance. That is to say, LR model trends to recommend the popular apps in top positions and is more likely to ignore the specific interest of different users. In contrast, due to the superior ability on capturing feature interactions, DeepFM-D is able to capture specific interests of different users better.

\section{Conclusions}\label{section:conclusion}

In this paper, we proposed DeepFM, an end-to-end wide \& deep learning framework for CTR prediction, to overcome the shortcomings of the state-of-the-art models. DeepFM trains a deep component and an FM component jointly. It gains performance improvement from these advantages: 1) it does not need any pre-training; 2) it learns both high- and low-order feature interactions; 3) it introduces a sharing strategy of feature embedding to avoid feature engineering. We studied two instances of DeepFM framework, namely DeepFM-D and DeepFM-P, of which the deep component are DNN and PNN, respectively. The offline experiments on three real-world data sets demonstrate that 1) our proposed DeepFM-D and DeepFM-P outperform the state-of-the-art models in terms of AUC and Logloss on all the three datasets; 3) As one of the best performed model, DeepFM-D has comparable efficiency as LR model on GPU, which is acceptable in industrial applications.

To verify the superiority of DeepFM framework in production environment, we deployed DeepFM-D in the recommendation engine of Huawei App Market. We also covered related practice in deploying our framework, such as multi-GPU architecture and asynchronous data reading.
Compared with a well-engineered LR model, which is one of the most popular CTR prediction models, DeepFM-D achieves more than 10\% improvement of CTR in online A/B test.




\ifCLASSOPTIONcaptionsoff
  \newpage
\fi



\bibliographystyle{IEEEtran}
\bibliography{complete}
%


%

\vspace{-15 mm}
\begin{IEEEbiography}[{\includegraphics[width=1in,height=1.1in,clip,keepaspectratio]{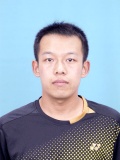}}]{Huifeng Guo} received the B.S. and M.S. degrees in Computer Science from Lanzhou University and Harbin Institute of Technology in China in 2012 and 2014, respectively. He is currently working toward the Ph.D. degree in the department of Computer Science, Shenzhen Graduate School, Harbin Institute of Technology. His research interests are in the areas of machine learning and recommendation, including graph mining and deep learning.
\end{IEEEbiography}
\vspace{-15 mm}
\begin{IEEEbiography}[{\includegraphics[width=1in,height=1.1in,clip,keepaspectratio]{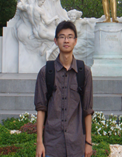}}]{Ruiming Tang} is currently a Senior Researcher in Huawei Noah's Ark Lab. He received the B.S. degrees from the Department of Computer Science at Northeastern University China in 2009, and the Ph.D. degree from the Department of Computer Science at National University of Singapore, in 2014. He joined Huawei Noah's Ark Lab since 2014. His research interests include machine learning and artificial intelligence, particularly, deep learning and recommender systems.
\end{IEEEbiography}
\vspace{-15 mm}
\begin{IEEEbiography}[{\includegraphics[width=1in,height=1.1in,clip,keepaspectratio]{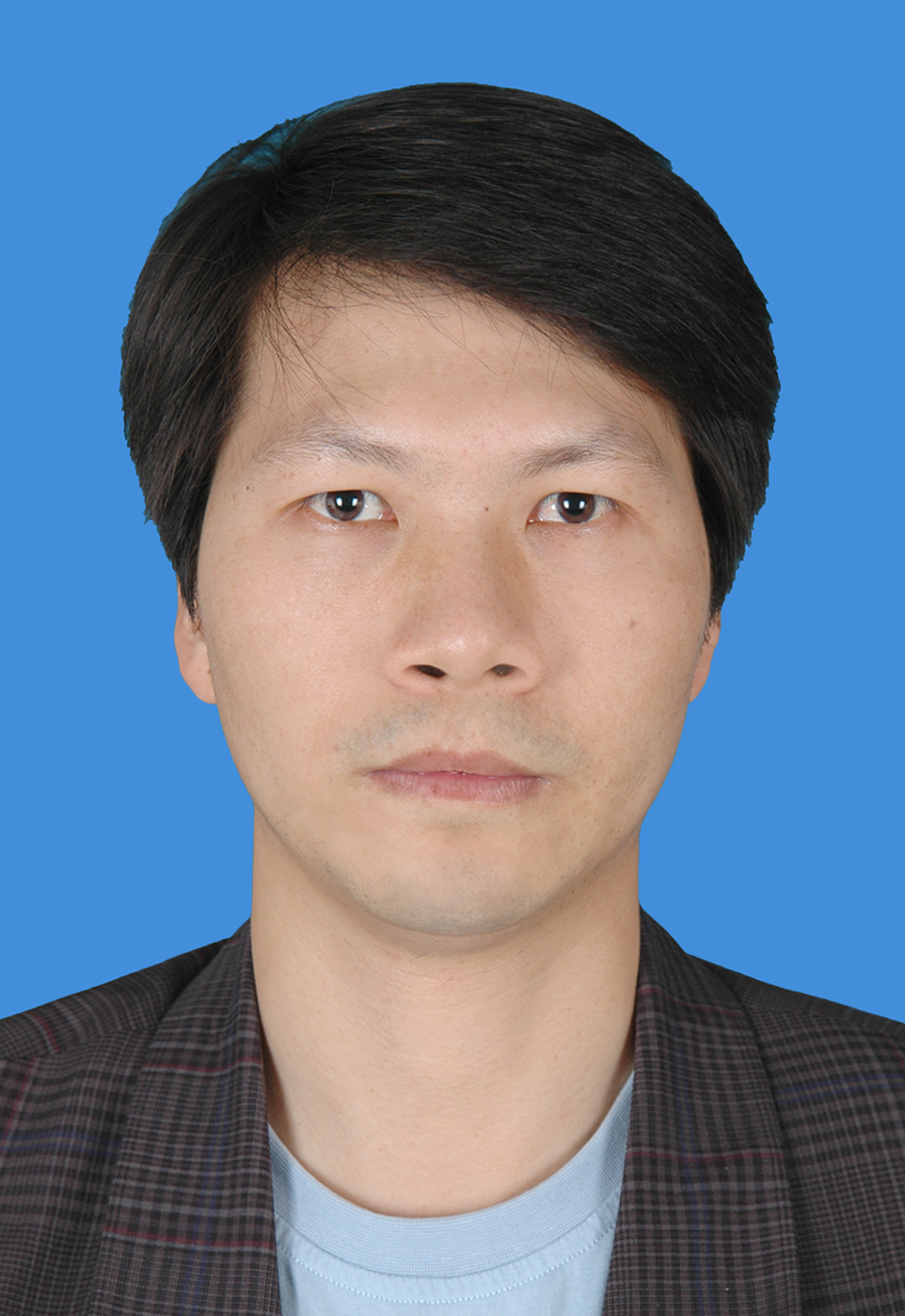}}]{Yunming Ye} received the Ph.D. degree in Computer Science from Shanghai Jiao Tong University. He is now a professor in the Shenzhen Graduate School, Harbin Institute of Technology. His research interests include data mining, text mining, and ensemble learning algorithms.
\end{IEEEbiography}
\vspace{-15 mm}
\begin{IEEEbiography}[{\includegraphics[width=1in,height=1.1in,clip,keepaspectratio]{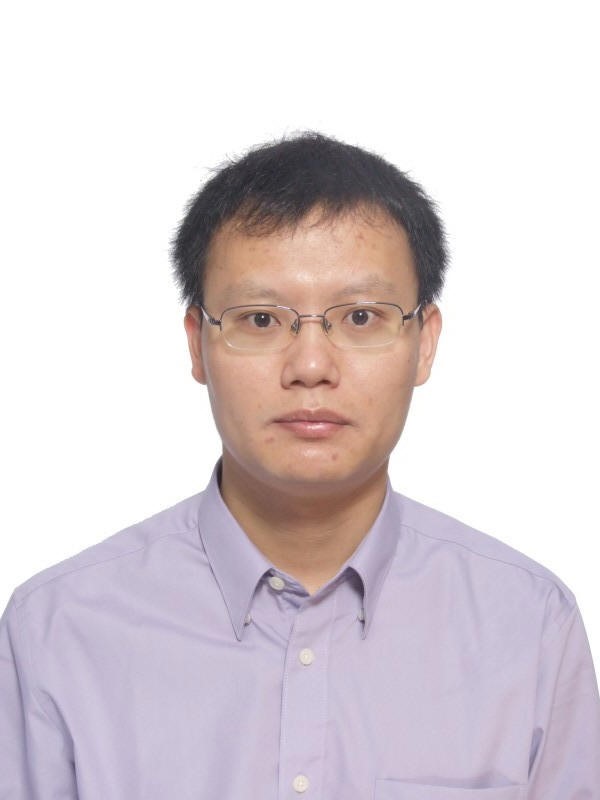}}]{Zhenguo Li} is currently the director of AI Theory Lab and Principal Researcher in Huawei Noah's Ark Lab. He received the B.S. and M.S. degrees from the Department of Mathematics at Peking University, in 2002
and 2005, respectively, and the Ph.D. degree from the Department of Information Engineering at the Chinese University of Hong Kong, in 2008. He was an associate research scientist in the Department of Electrical Engineering at Columbia University. His research interests include machine learning and artificial intelligence.
\end{IEEEbiography}
\vspace{-15 mm}
\begin{IEEEbiography}[{\includegraphics[width=1in,height=1.1in,clip,keepaspectratio]{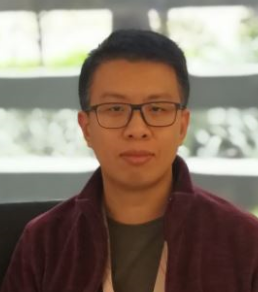}}]{Xiuqiang He} is currently the director of Recommendation \& Search Lab and Principal Researcher in Huawei Noah's Ark Lab. He received the B.S. and M.S. degrees from the Department of Computer Science at Xi'an Jiaotong University, in 2003 and 2006, respectively, and the Ph.D. degree from the Department of Computer Science at the Hong Kong University of Science and Technology, in 2010. His research interests include machine learning algorithms in the area of recommendation and search.
\end{IEEEbiography}
\vspace{-15 mm}
\begin{IEEEbiography}[{\includegraphics[width=1in,height=1.1in,clip,keepaspectratio]{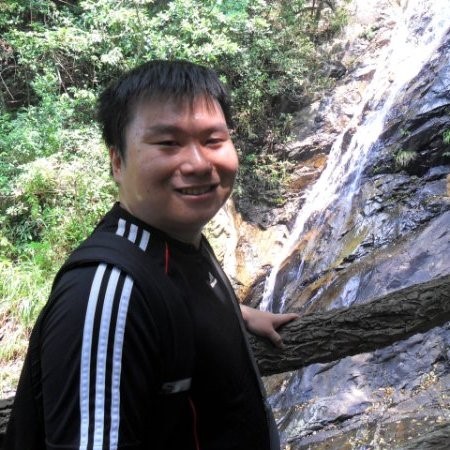}}]{Zhenhua Dong} is a Senior Researcher of Huawei Noah's ark lab, he received the BEng degree from Tianjin University in 2006 and the PhD degree in computer science and technology from Nankai University, China, in 2012. He was a research assistant at GroupLens lab in the University of Minnesota during 2010-2011. His research interests include recommender systems, mobile computing.
\end{IEEEbiography}
\end{document}